\documentstyle[graphicx]{mn}

\include{graphicx}

\title[NGC\,4395]
{Optical and X-ray Variability in The Least Luminous AGN, NGC\,4395}

\author[Lira et al.]
{P.~Lira,$^1$ A.~Lawrence,$^1$ P.~O'Brien,$^2$ R.~A.~Johnson,$^3$ 
R.~Terlevich$^4$\thanks{Visiting Professor at the Instituto Nacional de Astrof\'\i sica, Optica y Electr\'onica, A.P. 51 y 216. 71200 Puebla; M\'exico.} 
\& N.~Bannister$^2$ \\
$^1$ Institute for Astronomy, University of Edinburgh, Royal Observatory, 
Blackford Hill, Edinburgh EH9 3HJ, Scotland\\
$^2$ Department of Physics \& Astronomy, University of Leicester, 
University Road, Leicester, LE1 7RH, UK\\
$^3$ Institute of Astronomy, Madingley Road, Cambridge CB3 0HA, UK\\
$^4$ Royal Greenwich Observatory, Madingley Road, Cambridge CB3 0EZ, UK}

\begin{document}

\maketitle

\begin{abstract}

We report the detection of optical and X-ray variability in the least
luminous known Seyfert galaxy, NGC\,4395. Between July 1996 and
January 1997 the featureless continuum changed by a factor of 2, which
is typical of more luminous AGN. The largest variation was seen at
shorter wavelengths, so that the spectrum becomes `harder' during
higher activity states. During the same period the broad emission-line
flux changed by $\sim 20-30$ per cent. In a one week optical broad
band monitoring program, a 20 per cent change was seen between
successive nights. The difference in flux observed between the
spectroscopy of July 1996 and the broad-band observations implies
variation by a factor of 3 at 4400 \AA\ in just one month. In the same
period, the spectral shape changed from a power law with spectral
index $\alpha \sim 0$ (characteristic of quasars) to a spectral index
$\alpha \sim 2$ (as observed in other dwarf AGN). {\it ROSAT\/} HRI
and PSPC archive data show a variable X-ray source coincident with the
galactic nucleus. A change in X-ray flux by a factor $\sim 2$ in 15
days has been observed. When compared with more luminous AGN,
NGC\,4395 appears to be very X-ray quiet. The hardness ratio obtained
from the PSPC data suggests that the spectrum could be absorbed. We
also report the discovery of weak CaIIK absorption, suggesting the
presence of a young stellar cluster providing of the order of 10\% of
the blue light. The stellar cluster may be directly observed as a
diffuse component in {\it HST\/} optical imaging.  Using {\it HST\/}
UV archive data, together with the optical and X-ray observations, we
examine the spectral energy distribution for NGC\,4395 and discuss the
physical conditions implied by the nuclear activity under the standard
AGN model. When in the low state, the extrapolated UV continuum is
insufficient to explain the observed broad emission-lines. This could
be explained by intrinsic variability or absorption or may imply an
extra heating source for the BLR. The observations can be explained by
either an accreting massive black hole emitting at about $10^{-3}$
$L_{Edd}$ or by a single old compact SNR with an age of 50 to 500
years generated by a small nuclear starburst.

\end{abstract}


\section{Introduction}

\begin{table*}
\centering
\caption{Journal of Observations of NGC\,4395}
\begin{tabular}{r@{\hspace{1cm}}c@{\hspace{1cm}}c@{\hspace{1cm}}c} \hline
Date & Telescope & Mode & Archive Data \\ 
&&& \\ 
7 April 1988 	& Hale & Spectroscopy & No$^{\dag}$ \\
2 July 1992 	& {\it ROSAT\/} PSPC & Imaging & Yes \\ 
15 \& 19 July 1992 & {\it HST\/} FOS & Spectroscopy & Yes \\
17 July 1992 	& {\it ROSAT\/} PSPC & Imaging & Yes \\ 
5 December 1995 & {\it HST\/} WFPC2 & Imaging & Yes \\ 
5 -- 16 June 1996 & JKT & Imaging & No \\ 
23 June 1996 	& {\it ROSAT\/} HRI & Imaging & No \\ 
13 July 1996 	& WHT & Spectroscopy & No \\ 
15 January 1997 & WHT & Spectroscopy & No \\ \hline
\multicolumn{4}{l}{$^{\dag}$ Data provided by Filippenko et al. See Filippenko \& Sargent 1989.}
\end{tabular}
\end{table*}

The dwarf Seyfert nucleus in NGC\,4395 was first reported by
Filippenko \& Sargent (1989) almost a decade ago. Optical spectroscopy
showed high ionization narrow lines as well as broad permitted
emission-lines. The detection of a compact radio source (Sramek 1992)
added support to the idea that NGC\,4395 is a feeble version of the
more luminous Seyfert galaxies. Its low luminosity nucleus has a blue
absolute magnitude $M_{B} \sim -11$, a luminosity $10^{4}$ times
fainter than a classical Seyfert galaxy like NGC\,4151. The detection
of a featureless UV continuum gave support to the idea that NGC\,4395
was a real example of dwarf nuclear activity (Filippenko, Ho \&
Sargent 1993). However, Shields \& Filippenko (1992), after several
years of spectroscopic monitoring, reported that no evidence was found
for continuum or line variability. Since variability is one of the
most common characteristics of AGN, this result was quite surprising.

NGC\,4395 is a nearly face-on dwarf galaxy ($B \sim 10.7$, $M_{B} \sim
-17.9$, assuming a distance of 5.21 Mpc - see below) with
morphological classification Sd III-IV in the extended Hubble system,
as defined by Sandage \& Tammann in the RSA catalogue (Sandage \&
Tammann 1981). It exhibits a star-like nucleus and an extremely low
surface-brightness disk. The loose and disconnected spiral arms show
some blue knots of star formation activity (for a colour plate see Wray
1988).

We will use a distance to NGC\,4395 of 5.21 Mpc, based on the observed
recession velocity and a Virgo flow model with an infall velocity of
220 km s$^{-1}$ for the Galaxy (Kraan-Korteweg 1986). This value
agrees well with the distance modulus of 28.5 reported by Sandage \&
Bedke (1994) and gives a scale of 25 parsecs per arcsec on the sky.

As part of a multiwavelength imaging and spectroscopy project to study
a volume-limited sample of very nearby galaxies we have acquired {\it
ROSAT\/} HRI data and optical spectroscopy for NGC\,4395. $B$ and $I$
images of the galaxy were obtained as part of an AGN monitoring
program. Finally, {\it HST\/} and {\it ROSAT\/} PSPC data for
NGC\,4395 were retrieved from public archives. In this paper we report
the detection of optical and X-ray variability in the nucleus of
NGC\,4395. In Section 2 we present the observations and reduction of
the X-ray imaging, optical spectroscopy, and broad-band optical data.
Results are given in section 3. A discussion is presented in Section
4, and the conclusions are summarized in Section 5.

\section{Observations and Data Reduction}

A journal with the data used in this paper can be found in table 1. It
includes ground based and {\it HST\/} imaging and spectroscopy,, and
{\it ROSAT\/} data. This section will describe the observations and
reduction of data which do not come from public archives.

\subsection{HRI data}

\begin{figure*}
\centering
\includegraphics[scale=0.9,bb=40 200 515 655]{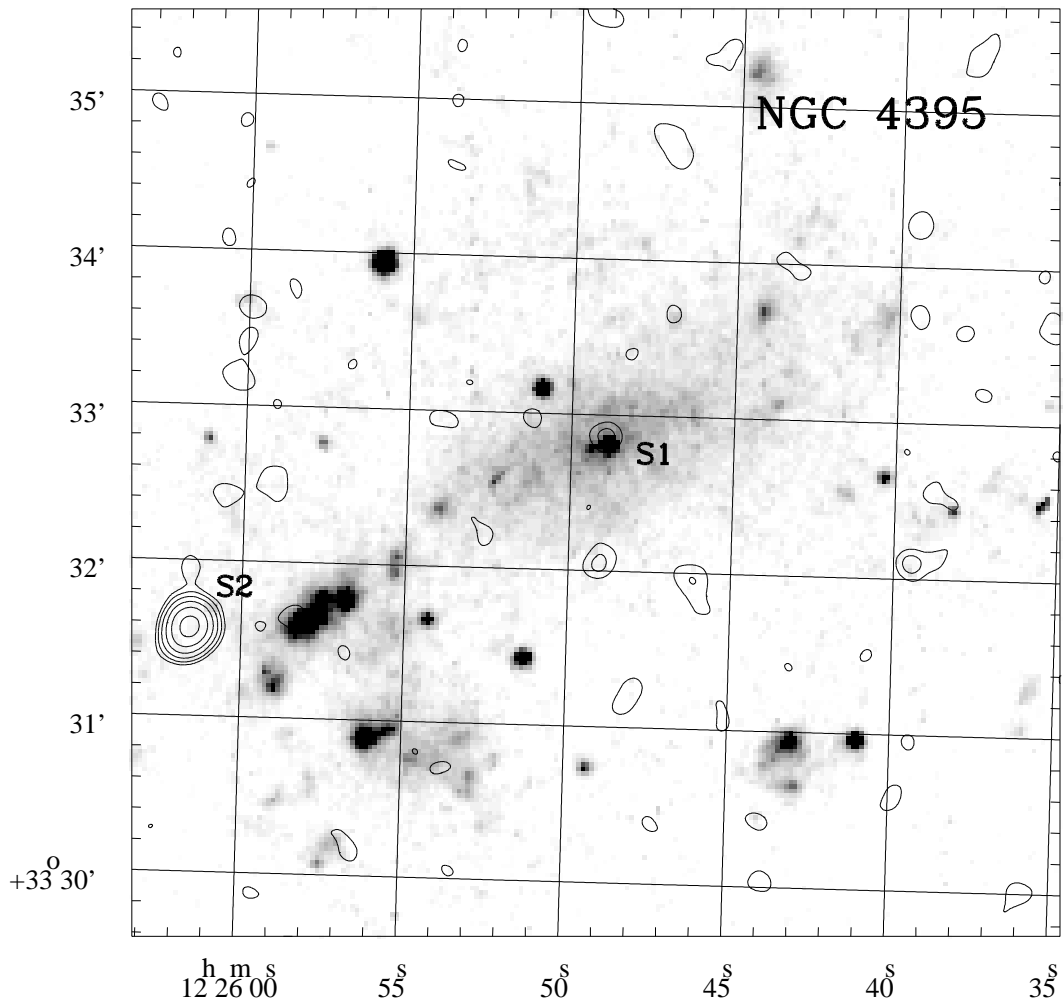}
\caption{{\it ROSAT\/} HRI contour plot overlaid onto a DSS plate of
NGC\,4395. Contours were drawn at 3, 6, 12, 24, 48 and 96 times the
standard deviation in the smoothed backgroud. S1 corresponds to
NGC\,4395 nucleus. S2 is a strong X-ray source without obvious optical
identification.}
\end{figure*}

An X-ray image of NGC\,4395 was obtained with the High Resolution
Imager (HRI) on board {\it ROSAT\/} on the 23rd of June 1996 as part
of the AO7 cycle of pointed observations. The total exposure time was
11,253 secs.

Figure 1 shows a contour map of the central part of the X-ray frame
overlaid on the Digital Sky Survey data for NGC\,4395. The pixels of
the HRI frame were binned to $2 \times 2$ arcsec$^{2}$ and the image
was convolved with a Gaussian of $\sigma =$ 4 arcsec. In this way the
noise in the image is artificially suppressed and can be used to draw
the contour levels. Contours in figure 1 were drawn at 3, 6, 12, 24,
48 and 96 times the standard deviation in the smoothed background
(count rates, fluxes and related errors elsewhere in this paper are
determined from the raw data).  The most prominent source in the
figure (S2) has no obvious optical identification although the DSS
image shows some diffuse emission in the area. We have identified the
possible source S1 as the tentative X-ray nuclear emission of
NGC\,4395. The slight shift between the optical nucleus and the X-ray
emission ($\sim 3.5$ arcsec) is consistent with the accuracy of the
{\it ROSAT\/} pointing. Indeed, another X-ray source approximately 7
arcmin to the west of S1 and with a clear optical identification shows
the same shift.

The count rate for the nucleus was calculated by summing over all the
counts within a circle centered on the source. To estimate the
background two circles of radius 160 arcsec free of evident X-ray
sources and away from the galaxy were used. To find the optimum radius
for the aperture radial profiles of several point sources were
examined. A final aperture of 10 arcsec was adopted which should
encircle $\sim$ 99 per cent of the photons at 0.2 keV and $\sim$ 86
per cent of the photons at 1.7 keV (David et al.\ 1997). The net count
for S1 was 7.6 $\pm$ 4.6 photons, i.e., it is not a significant
detection. For S2 we find a net count of 169.2 $\pm$ 14.3 photons. The
count rates are $6.6 \pm 4.0 \times 10^{-4}$ and $15.0 \pm 1.3 \times
10^{-3}$ photons s$^{-1}$ for S1 and S2 respectively.

\subsection{PSPC data}

Two sets of PSPC data were retrieved from the {\it ROSAT\/}
archive. One of these data sets is presumably that referred to as a
private communication from Snowden \& Belloni in Filippenko, Ho \&
Sargent (1993). The first set was obtained on the 2nd of July 1992
with 7,755 seconds of exposure time, while the second set was obtained
15 days later with 8,764 seconds exposure. Comparing both data sets it
is easy to identify a variable X-ray source which is consistent with
the position of the nuclear source for NGC\,4395 marginally detected
from our HRI image.

For total count extraction of a PSPC point source, an aperture of 2
arcmin should be adequate. Reducing the aperture to 1 arcmin loses
15-20 per cent of the counts at the soft end of the spectrum (E $\la$
0.1 keV) because of the wider PSF at lower energies (Hasinger et al.\
1992). However, apertures larger than 30 arcsec around the NGC\,4395
nuclear source would include other knots of X-ray emission, as can be
seen in our HRI image. To estimate the flux due to these extra-nuclear
sources, we measured the net counts in the HRI image using an annuli
centered on the nuclear source with an inner radius of 10 arcsec and
an outer radius of 1\arcmin. We find 23.5 $\pm$ 14.7 counts in the
annuli, so we expect some contamination within the 1 arcmin aperture,
but it should not be significant.

For a 1 arcmin aperture centered at the position of the NGC\,4395
nucleus, the net counts were $67.2 \pm 10.7$ photons and $139.7 \pm
14.1$ photons for the first and second PSPC data sets, respectively.
For the background estimation we used a large circle of radius 600
arcsec far away from any contamination by other X-ray sources. The
associated count rates are $8.7 \pm 1.4 \times 10^{-3}$ photons
s$^{-1}$ on the 2nd of July 1992, and $15.9 \pm 1.6 \times 10^{-3}$
photons s$^{-1}$ 15 days later, giving a variability of about a factor
of two.

\subsection{Optical spectroscopy}

\begin{figure*}
\centering
\includegraphics[scale=0.9,bb=30 450 580 700]{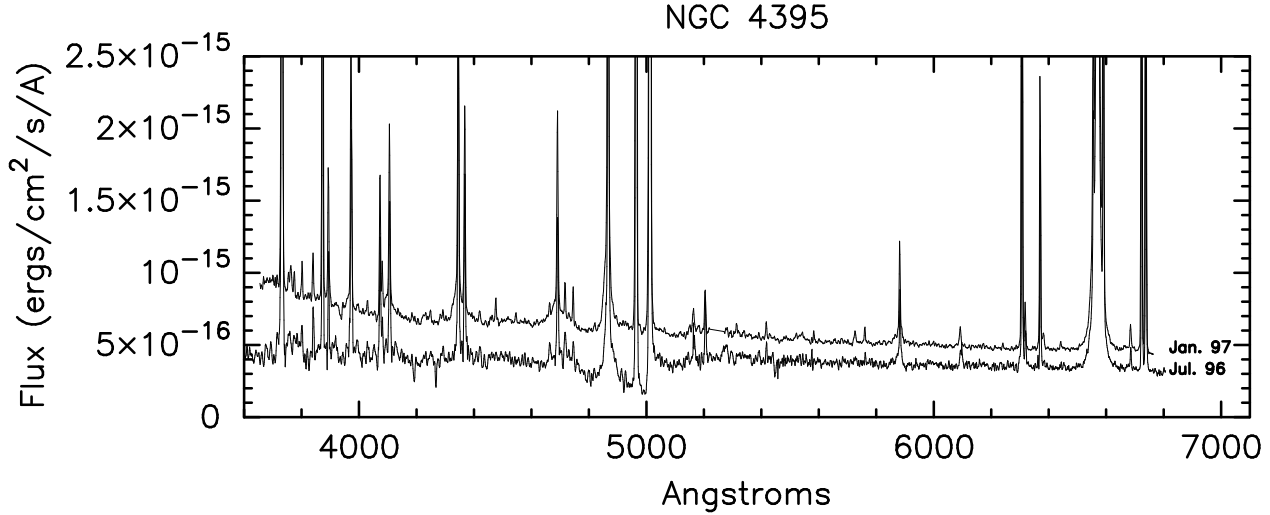}
\caption{Optical spectra obtained with the WHT showing high and low
state of NGC\,4395. The lower spectrum was obtained on July 1996 and
the upper was obtained almost exactly 6 months later. For plotting
purposes, the blue continuum from the July 1996 spectrum has been
slightly smoothed in order to suppress noise. Fluxes are in units of
ergs cm$^{-2}$ s$^{-1}$ \AA$^{-1}$.}
\end{figure*}

Long slit spectra of the nucleus of NGC\,4395 were obtained on the
13th of July 1996 and the 15th of January 1997 at the Cassegrain focus
of the 4.2\,m William Herschel Telescope at the Roque de los Muchachos
Observatory. The R316R grating was installed in the red arm of the
ISIS double spectrograph and the R300B was used in the blue arm,
together with dichroic 5400 (half power point of crossover at 5470
\AA). The wavelength coverage was $\sim$ 3650 - 6750 \AA. A small gap
between 5220 and 5270 \AA\ was not observed during January 1997. A TEK
CCD and a Loral CCD were used during July 1996 for the red and the
blue camera, respectively. Both cameras were equipped with TEK CCDs
for the run in January 1997. The slit width was 1 arcsec on the sky,
resulting in a spectral resolution of $\sim$ 3.0 \AA\ FWHM for the red
arm and $\sim$ 3.5 \AA\ FWHM for the blue arm ($\sim$ 3.8 \AA\ FWHM
with the Loral chip). The seeing varied between $\sim 0.8$ and $\sim
1.2$ arcsecs during both runs. The dispersion achieved in the red arm
was 1.47 \AA\ per pixel (0.061 \AA\ per micron). In the blue arm the
dispersion was 0.96 \AA\ per pixel with the Loral and 1.54 \AA\ per
pixel with the Tek detector (0.065 and 0.064 \AA\ per micron,
respectively). In both runs the slit was positioned at the parallactic
angle to minimize light losses. The CCDs were windowed to cover 4
arcmin in the spatial direction with a scale of 0.36 arcsec per pixel
for the red arm and 0.20 arcsec per pixel for the blue arm. In July
1996 conditions were photometric during the whole night, although the
presence of Saharan dust in the atmosphere hampered some of the
observations. The conditions during January 1997 were not photometric
throughout the night, but the data for NGC\,4395 and associated
standard stars were acquired during clear periods of the night.

\begin{table*}
\centering
\caption{Narrow Line Fluxes}
\begin{tabular}{lccccccccccc} \hline
& \multicolumn{4}{c}{ Blue Arm } & & \multicolumn{6}{c}{ Red Arm } \\ 
Date & [OII] & [NeIII] & [OIII] & [OIII] & & [OI] & [OI] & [NII] & [NII] & [SII] & [SII] \\ 
& $\lambda$3727 & $\lambda$3869 & $\lambda$4959 & $\lambda$5007 & & $\lambda$6300 & $\lambda$6363 & $\lambda$6548 & $\lambda$6583 & $\lambda$6716 & $\lambda$6731 \\ 
April 1988 $^{\dag}$ & -- & -- & 64.41 & -- && 23.02 & 7.59 & -- & -- & 19.77 & 22.97 \\ 
July 1996 & 36.83 & 19.19 & 66.84 & 207.60 && 19.60 & 6.52 & 4.29 & 12.86 & 13.00 & 16.57 \\
January 1997 & 36.34 & 21.55 & 69.72 & 227.82 && 24.69 & 7.77 & 4.80 & 14.40 & 16.34 & 19.81 \\ 
\hline
\multicolumn{12}{l}{Fluxes in units of 10$^{-15}$ ergs s$^{-1}$ cm$^{-2}$} \\
\multicolumn{12}{l}{$\dag$: Line fluxes from the spectrum acquired by Filippenko et al.}
\end{tabular}
\end{table*}

\begin{table*}
\centering
\caption{Broad and Narrow Deblended Line Fluxes}
\begin{tabular}{lccccccccccccc} \hline
Date & H$\beta_{N}$ & FWHM && H$\beta_{B}$ & FWHM && H$\alpha_{N}$ & FWHM && H$\alpha_{B}$ & FWHM$_{1}$ & FWHM$_{2}$ \\
April 1988 $^{\dag}$ 	& 28.58 & 4.0 && 17.43 & 21.3 &&  --   & --  &&   --   & -- & -- & \\ 
July 1996 		& 21.80 & 4.0 && 14.29 & 22.4 && 55.20 & 3.5 && 82.90  & 10.0 & 36.4 \\
January 1997 		& 24.20 & 4.0 && 18.49 & 20.4 && 59.85 & 3.6 && 120.86 & 9.7  & 33.9 \\
\hline
\multicolumn{13}{l}{Fluxes in units of 10$^{-15}$ ergs s$^{-1}$ cm$^{-2}$} \\
\multicolumn{13}{l}{$\dag$: Line fluxes from the spectrum acquired by Filippenko and co.}
\end{tabular}
\end{table*}

\begin{figure*}
\centering
\includegraphics[scale=0.8,bb=30 450 580 700]{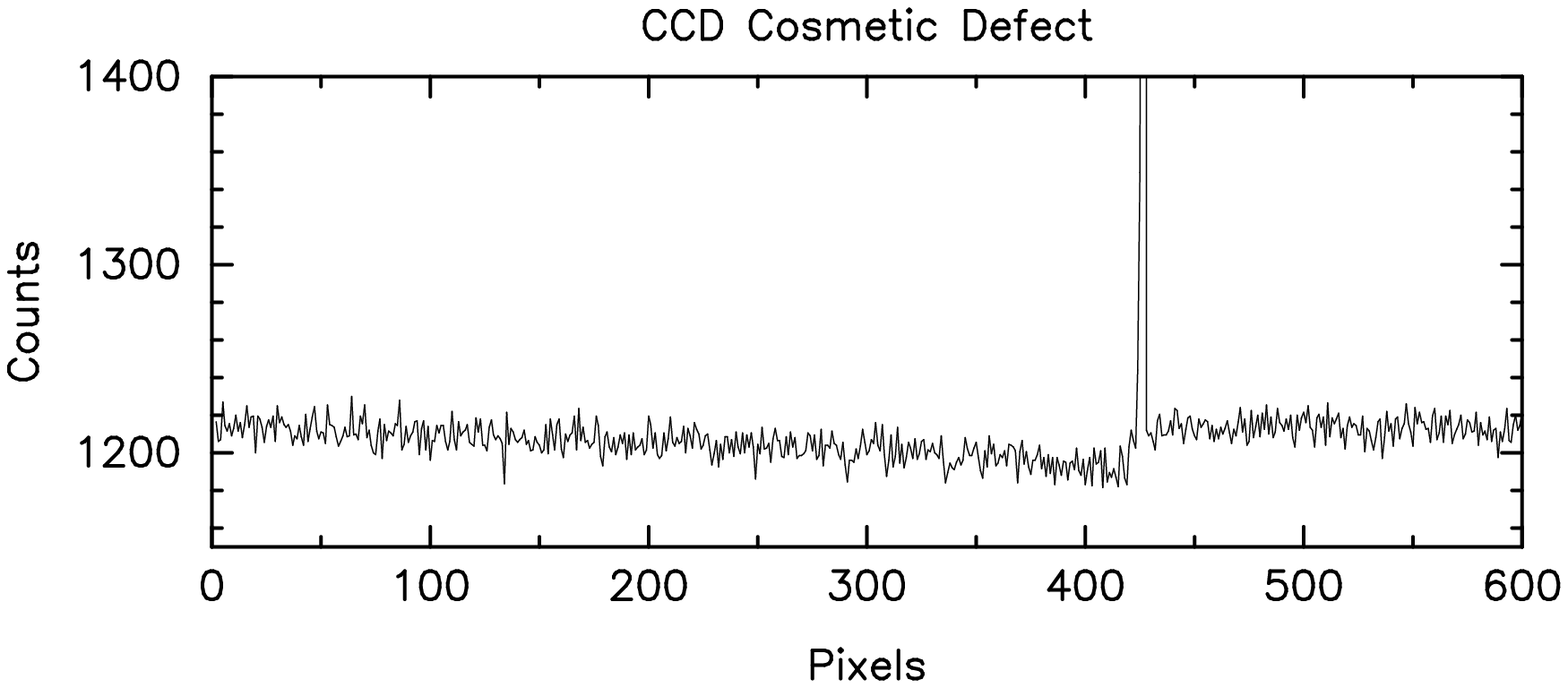}
\caption{Dark tail produced after bright pixels in a raw frame
obtained with the Loral CCD.}
\end{figure*}

The data were reduced using {\sc iraf} software. Bias correction and
flat-fielding of the 2D frames were performed in the usual way.
Spectra of the nucleus were extracted using an effective aperture of
$\approx 1 \times 2$ arcsec$^{2}$. For wavelength calibration a fifth
order Legendre polynomial was fitted to the strongest non-blended
emission-lines of copper-neon and copper-argon lamps. The frames were
flux calibrated using the spectrophotometric standards BD+28\,4211 and
PG1708+602 in July 1996, and Feige\,34 and G191-B2B in January 1997,
and using the mean extinction curve for the Observatory
($E_{\lambda}$). The extinction correction is air-mass dependent and
can be expressed as $A \times E_{\lambda}$ (in magnitudes), where $A$
is the air-mass during the observations. For the 13th of July 1996 a
grey shift correction of 0.13 magnitudes< (as measured on site by the
Carlsberg Meridian Circle) was applied to the mean extinction curve to
account for the dust extinction. This is possible because the
wind-blown Saharan dust in the atmosphere above the Canary Islands is
grey to an accuracy $\la$ 5 per cent between 0.32 and 1.0 microns
(Stickland et al.\ 1987; Whittet, Bode \& Murdin 1987). For January
1997 the flux calibration was done using the standard extinction curve
with no other corrections. As the standard stars were observed with a
wider slit (8 arcsecs), an empirical aperture correction was applied
to the NGC\,4395 spectra. The correction was calculated by comparing
the standard star spectra used in the flux calibration with spectra of
the same stars obtained by software which simulated a narrow slit. A
slight variation with wavelength was found and the NGC\,4395 spectra
were scaled by a factor of $\sim 1.4$ at the blue end ($\lambda \sim
3700$ \AA) and by a factor of $\sim 1.3$ at the red end ($\lambda \sim
6800$ \AA). The resulting spectra are shown in figure 2. The blue
continuum from the July 1996 spectrum has been smoothed slightly so
that its overall shape is easier to follow.

The interstellar Galactic absorption towards NGC\,4395 is just $A_{V}
= 0.008$ mag (Burstein \& Heiles 1984), and no correction to account
for this extinction was used in the reduction of the spectra. The
signal to noise per pixel achieved during July 1996 was $\sim$ 16 for
the red arm and $\sim$ 8 per for the blue arm. For January 1997 the
signal to noise per pixel was $\sim$ 42 and $\sim$ 35 for the red and
blue arm respectively.

{\it HST\/} Planetary Camera observations of the NGC\,4395 taken in
the narrow band F502N filter show that the [OIII]\,$\lambda$5007
emission region has a diameter of $\la$ 0.4 arcsec (Filippenko, Ho \&
Sargent 1993). In contrast, the FWHM for the spatial profile of an
unresolved star in our data is $\ga$ 1 arcsec.  Inspection of the
profile of the emission-lines in our spectra shows that the Narrow
Line Region (NLR) is, as expected, spatially unresolved and no
aperture effects have to be taken into account when comparing narrow
line fluxes from our two spectra.

Measurements of the narrow line fluxes in July 1996 and January 1997
agree to within 25 per cent at worst (see table 2). Since we have used
a 1 arcsec slit, the accuracy of the absolute calibration is expected
to be of the order of 30 per cent.  Comparison of the narrow line
fluxes measured from our data with the values measured from data
obtained in 1988, and kindly provided by Filippenko and collaborators,
have also been included in table 2 and 3. The spectra were examined
using the same software packages used in the reduction of our
data. The line fluxes agree to within 30 per cent, except for
[SII]\,$\lambda$6716, where the difference is slightly bigger.

The blue spectrum obtained during the July 1996 observation shows a
depression just after the [OIII]\,$\lambda$5007 emission-line. This is
due to a cosmetic defect in the CCD which produces a dark tail after
bright pixels, as can be seen in figure 3. The problem is only
apparent whenever pixels have $\ga$ 2000 counts in the raw frame. The
depth of the depression seems to be correlated with the peak count.
The data shown in figure 3 were produced with the largest cosmic ray
available in the image (49447 peak counts) and the depression extends
over up to 300 pixels ($\sim$ 270 \AA). Peak counts for the NGC\,4395
emission-lines are larger than 2000 counts only for the [OIII] lines,
making any depressions negligible for wavelengths shorter than $\sim$
5200 \AA. Unfortunately, fluxes for H$\beta$ could be slightly
affected.

\subsection{Ground-based broad-band observations}

As part of an AGN monitoring program, NGC\,4395 was observed on
5th--11th of June 1996 with the Jacobus Kapteyn Telescope
(JKT). Observations of NGC\,4395 were obtained over a period of a few
hours at the start of each night using the JKT CCD camera, using $B$
and $I$ band filters. An integration time of 5 minutes was used for
each exposure. The atmospheric conditions were judged to be good and
very stable over the entire week. This was confirmed by inspection of
the extinction data recorded independently by the Carlsberg automatic
Meridian Circle, which showed the average V-band extinction at the
zenith was constant to $\pm 0.01$ over the observing run.

The CCD data were reduced in a standard way using the {\sc iraf}
packages.  Although photometric standard stars were observed during
the monitoring campaign, we restricted our variability analysis to
photometry of the nucleus of NGC\,4395 relative to several stars
within the same CCD frames. This procedure is better suited to
searching for rapid, small-amplitude variability for which atmospheric
changes can significantly affect the results (e.g., Done et al.\
1990).  Within each CCD frame several nearby stars of similar
magnitude to the AGN nucleus were identified, and their counts
calculated using a circular photometric aperture 6 arcsec in radius. A
large aperture was used to ensure all the point-source light was
enclosed allowing for possible small variations in seeing. For each
star the sky background was removed by subtracting a scaled average of
the counts in an annulus of inner and outer radius 9 and 11 arcseconds
respectively centered on the photometric aperture. A similar procedure
was used for the AGN, except the annulus lay on top of the galaxy
thereby permitting a first-order correction for the galaxy-light
within the photometric aperture. No other attempt was made to correct
for galaxy contamination. We note, however, that the central region of
the galaxy of NGC\,4395 is quite faint even relative to the
low-luminosity nucleus: the surface brightness of the galaxy in the
nucleus vicinity is $\sim 20.5$ mag arcsec$^{-2}$. The light
distribution is also flat in the spatial direction (i.e., no strong
central bulge appears present). Hence the adopted galaxy-subtraction
procedure appears quite adequate for obtaining a good measurement of
the intrinsic, nuclear-variability amplitude.

\subsection{{\it HST\/} WFPC2 observations}

NGC\,4395 was observed with the WFPC2 on board {\it HST\/} in the
F450W ($\sim B$) and F814W ($\sim I$) band filters on the 5th of
December 1995 as part of the GTO proposal 6232. These data were
retrieved from the {\it HST\/} data archive and analysed using {\sc
iraf}. Results from previous analysis of these observations can be
found in Matthews et al.\ (1996;1998). The nucleus of NGC\,4395 was
imaged on the PC chip which has a pixel size of 0.046 arcsec. Three
exposures of NGC\,4395 were taken in each filter. The F450W exposure
times were 1$\times$60s and 2$\times$400s and the F814W exposure times
were 1$\times$60s and 2$\times$300s. We have only used the short
exposure observations here as the nucleus was saturated in the long
exposures.

\section{Results}

\subsection{X-ray imaging and spectral analysis}

The conversion to fluxes of the HRI and PSPC count rates for the
NGC\,4395 nuclear source were done assuming a power law spectrum
($F_{\nu} \propto \nu^{-\alpha}$) with energy index $\alpha = 1$ and
1.5, and using the energy range 0.1 -- 2.4 keV. Adopting a Galactic
hydrogen column density of $1.31 \times 10^{20}$ cm$^{-2}$ (Stark et
al.\ 1992) the HRI count rate gives a flux $\sim 3.5 \times 10^{-14}$
ergs s$^{-1}$ cm$^{-2}$, as can be seen in table 4. For a distance of
5.21 Mpc this implies an X-ray nuclear luminosity for NGC\,4395 of
$1.1 \times 10^{38}$ ergs s$^{-1}$.

Since at least 100 counts are required to perform spectral analysis of
PSPC observations, it was not possible to fit the data using the 34
energy channels of the detector. Instead, the total counts (0.1--2.4
keV) for each data set were binned into a single channel to estimate
the fluxes. A power law spectrum with $\alpha = 1$ and 1.5, and a
hydrogen column density of $1.31 \times 10^{20}$ cm$^{-2}$ were
assumed. The results are shown in table 4. 

In section 3.2 we discuss the possibility of a young stellar cluster
in the nucleus of NGC\,4395. If this is present it could be the
dominant source of very soft X-rays. During the high state the count
rate in the soft 0.1-0.4 keV band was $4.7 \pm 1.0 \times 10^{-3}$
photons s$^{-1}$, contributing about 30 percent of the total flux
($8.7 \pm 1.4 \times 10^{-3}$ photons s$^{-1}$).  During the low state
the soft count rate drops proportionately, continuing to provide about
30 percent of the total flux.  The source is therefore variable,
implying that the soft flux is probably dominated by nuclear emission.

In order to obtain spectral information from the PSPC data, we have
used the hardness ratio (HR) technique, which gives an `X-ray colour'
for objects with few net counts (Hasinger 1992; Ciliegi et al.\
1997). By definition HR = (H-S)/(H+S), where S is the number of net
counts within the channels 11-42 ($\sim 0.11 - 0.43$ keV), and H is
the number of counts in the channels 51-201 ($\sim 0.51 - 2.02$
keV). Values of HR close to -1 indicate that the source has an
extremely soft spectrum, while values close to +1 show that the source
has a hard or heavily absorbed spectrum. For NGC\,4395 HR = 0.37 $\pm$
0.11, where the error was calculated as in Ciliegi et al.\
(1997). This value of HR, with a Galactic hydrogen column density,
implies a spectral index of $\alpha \sim 1$.  If NGC\,4395 is indeed a
typical Seyfert 1 we would expect $\alpha_{x} \ga 1.5$ (Laor et al.\
1997 ; Walter \& Fink 1993), which would require an additional
hydrogen column density of $\sim 3 \times 10^{20}$ cm$^{-2}$ to
explain the observed HR. The hardness ratio for NGC\,4395, therefore,
is consistent with a modest intrinsic absorption of the soft part of
the X-ray spectrum. 

\begin{table}
\caption{{\it ROSAT\/} (0.1--2.4 keV) Fluxes}
\begin{tabular}{lccccc} \hline
& & HRI & & \multicolumn{2}{c}{PSPC}\\ 
& & & & Low State & High State\\
$\alpha = 1.5$ && 0.37 && 1.11 & 2.06 \\
$\alpha = 1.0$ && 0.32 && 1.10 & 2.05 \\ \hline
\multicolumn{6}{l}{Fluxes in units of $10^{-13}$ ergs s$^{-1}$ cm$^{-2}$}
\end{tabular}
\end{table}

\begin{figure*}
\centering
\includegraphics[bb=60 410 570 645,scale=0.8]{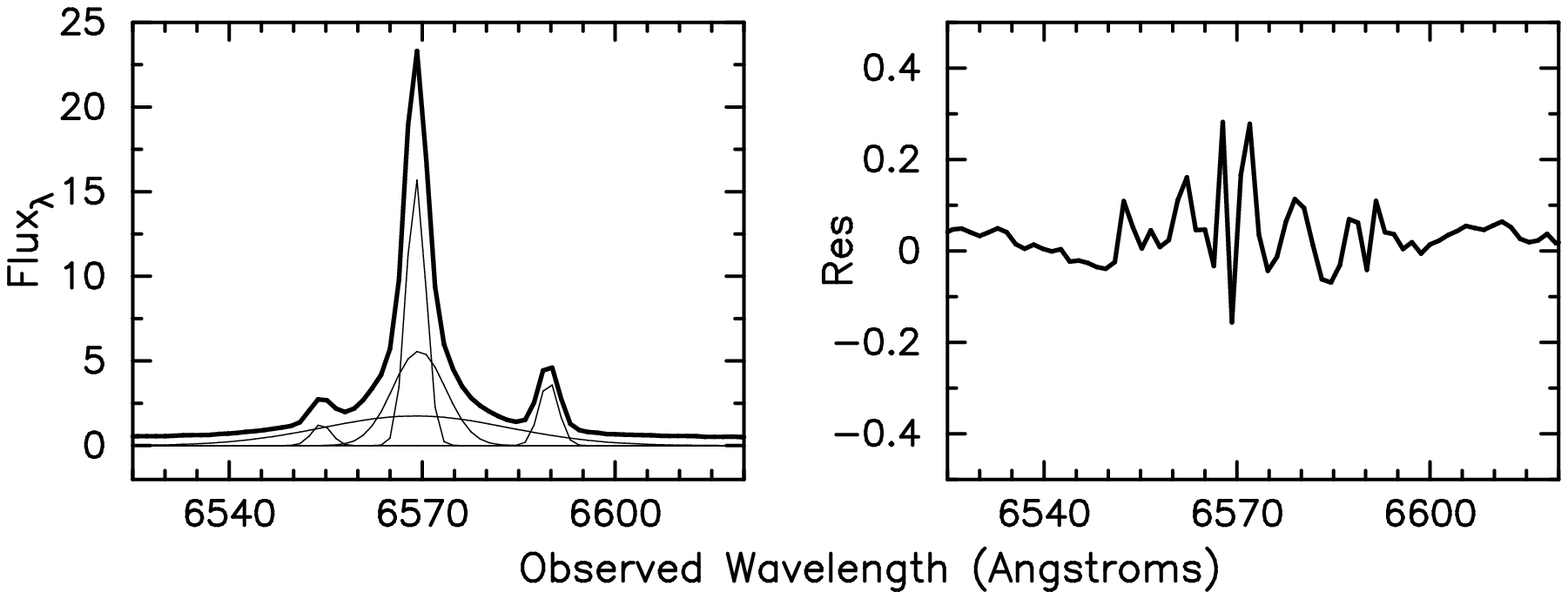}
\includegraphics[bb=60 410 570 645,scale=0.8]{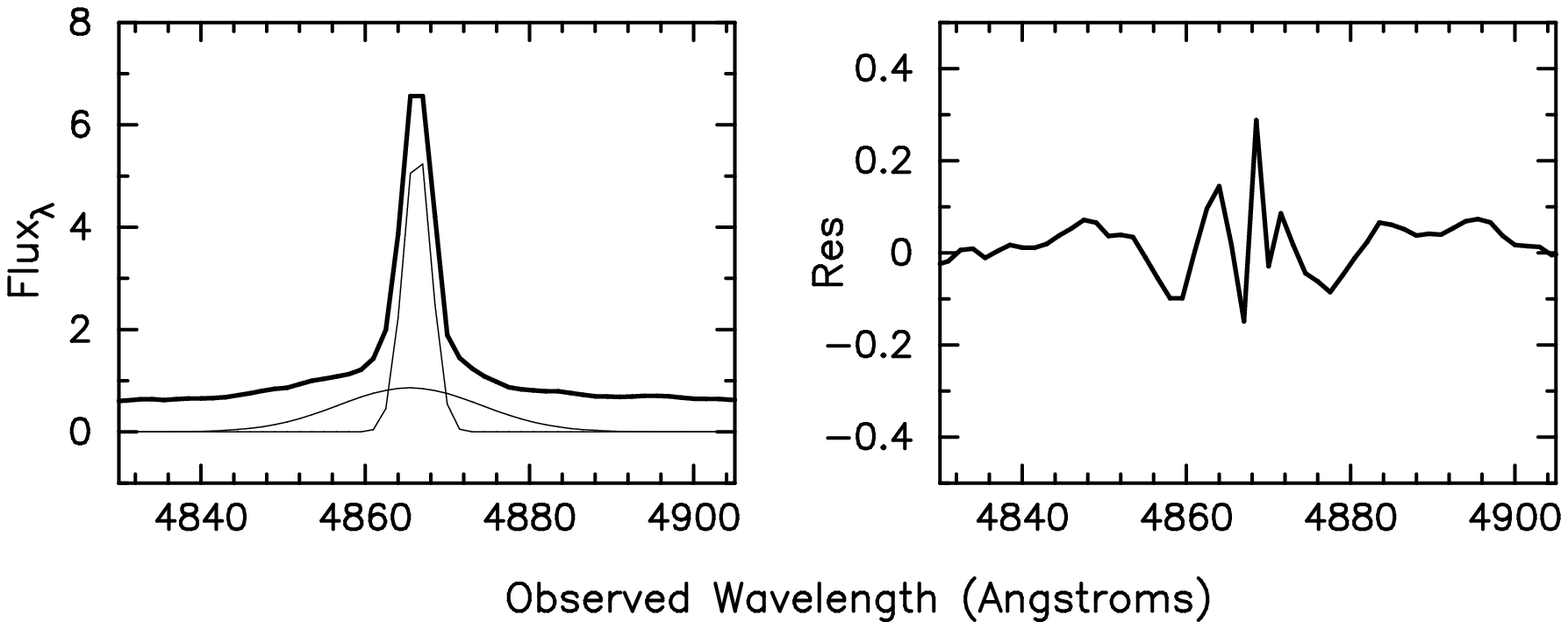}
\caption{Gaussian profile fitting to H$\alpha$ (top) and H$\beta$
(bottom). Two and one {\it broad\/} components were fitted to each
line, respectively. The narrow components had a fixed width measured
from other narrow lines. The left panel shows the data and individual
model components (the continuum level is not included). The right
panel shows the residuals between the data and model. All fluxes are
in units of $10^{-15}$ ergs cm$^{-2}$ s$^{-1}$ \AA$^{-1}$.}
\end{figure*}

\subsection{Analysis of optical spectra}

Figure 2 shows the spectra obtained in July 1996 and January
1997. Within 6 months the continuum has changed by a factor of $\sim$
1.3 at red end of the spectra and by a factor of $\sim$ 2.2 at the
blue end. The nuclear source becomes bluer when brighter, with a
change in the spectral index from $\alpha \sim 2$ to $\alpha \sim 1$
(see Section 4). From the narrow line fluxes quoted in table 2 it
seems that the flux at the red end of the spectrum obtained in January
1997 might be slightly overestimated when compared to the July 1996
observation. It is then possible that there is negligible change at
the red end of the spectra, and an even more dramatic change in colour
between July 1996 and January 1997. The continuum becomes harder when
it is brighter, which is a general characteristic of classical AGNs
(Kassebaum et al.\ 1997; Kaspi et al.\ 1996a; Reichert et al.\ 1994;
Peterson et al.\ 1991; Edelson, Krolik \& Pike 1990).

Two and three Gaussians were fitted to H$\beta$ and H$\alpha$
respectively: a narrow component with a fixed (instrumental) FWHM as
measured from the narrow lines, and one (or two) free parameter broad
components to fit the extended wings. The continuum level and slope
were also free parameters during the fitting. A Lorentzian profile was
also fitted to the broad components, but the results were much poorer
than when using Gaussian profiles. Figure 4 shows the fit to H$\alpha$
and H$\beta$ as individual Gaussian components and as residuals
between the data and model. Errors for the broad components were found
to be less than 3 percent. They were computed as the square root of
the diagonal elements of the covariance matrix of the non-linear model
(ie, they represent 68 percent confidence intervals for each parameter
taken separately).  Table 3 gives the fluxes and FWHM obtained for
each line. For H$\alpha$ the values for both fitted broad components
are shown.

\begin{figure*}
\centering
\includegraphics[scale=0.9,bb=60 320 500 660]{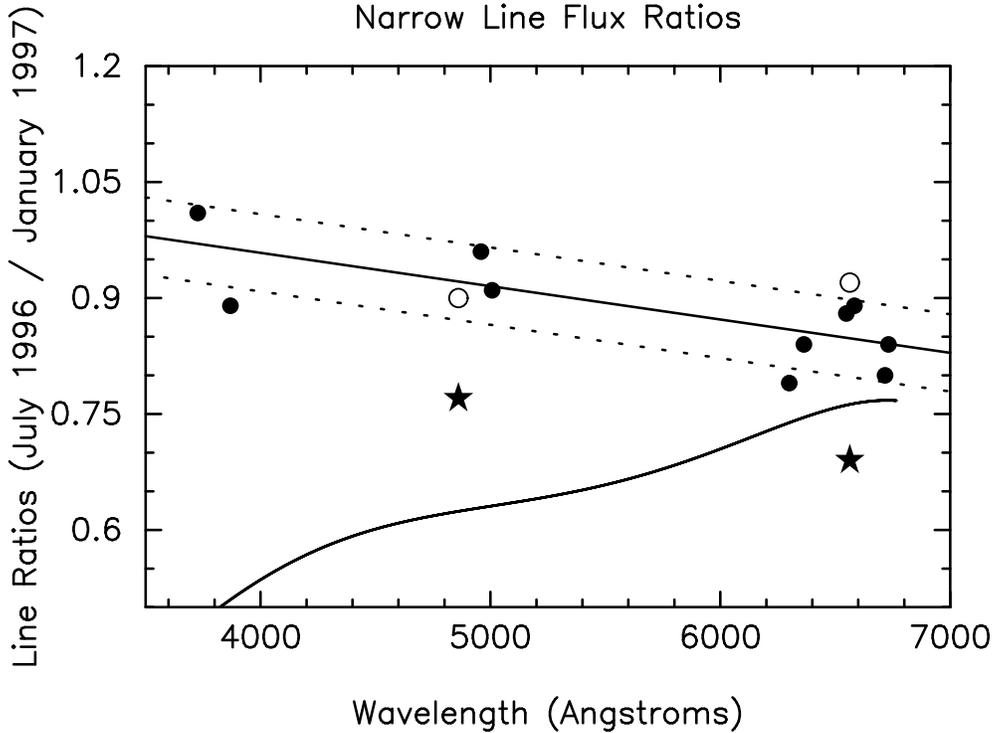}
\caption{Narrow line flux ratios between July 1996 and January 1997 as
a function of wavelength. A linear fit to the data is shown as a solid
line. Dashed lines correspond to 1 standard deviation from the
fit. The empty circles correspond to the ratio of the narrow line
component of H$\alpha$ and H$\beta$ which were not used in the
fit. The stars correspond to the ratio of the H$\alpha$ and H$\beta$
broad components. The ratio between the nuclear continuum observed in
July 1996 and in January 1997 is also shown as a curved line at the
bottom of the figure.}
\end{figure*}

The narrow line widths are not resolved in our data (resolution $\sim$
3.5 \AA). To find the fluxes given in table 2 the lines were fitted
with a single Gaussian (except for the nitrogen doublet). Gaussian
profile fitting to the Balmer lines determined the broad component
luminosities. Two and three Gaussians were fitted to H$\beta$ and
H$\alpha$ respectively: a narrow component with a fixed (instrumental)
FWHM as measured from the narrow lines, and one (or two) free
parameter broad components to fit the extended wings. The continuum
level and slope were also a free parameters during the fitting. Figure
4 shows the fit to H$\alpha$ and H$\beta$ as individual Gaussian
components and residuals. A Lorentzian profile was also fitted to
the broad components but the results were much poorer than when using
Gaussian profiles. Errors for the broad line fluxes were found to be
less than 3 percent. The errors were computed as the square root of
the diagonal elements of the covariance matrix for each parameter of
the non-linear model (ie, they represent 68 percent confidence
intervals for each parameter taken separately). Table 3 gives the
fluxes and FWHM obtained for each line. For H$\alpha$ the values for
both fitted broad components are shown.

The error on the absolute line fluxes is probably of the order of 30
per cent, but the {\it relative\/} changes in H$\alpha$ and H$\beta$
can be obtained to within a few percent by normalizing to nearby
narrow line fluxes. At a distance of 5.21 Mpc the NLR has a linear
size of $\sim 10$ pc (from {\it HST\/} PC narrow band observations
centred on the [OIII]\,$\lambda$5007 line; Filippenko, Ho \& Sargent
1993). Although fairly modest, this diameter implies a traveling time
of more than 30 years which should ensure that any variations in the
central continuum source will be smeared out within the NLR and that
the narrow line emission is fairly constant.

Given the observed size of the nuclear region (see sections 2.3 and
3.2) there should be no significant aperture effects to take into
account, so we can use the ratios of the narrow line fluxes observed
in July 1996 and January 1997 to assess any relative calibration
differences.  The observed narrow line ratios indicate that there is a
slight variation with wavelength in the relative calibration. By
fitting to these ratios, and assessing their scatter about the result,
we can estimate the significance of the broad line variations.

Figure 5 shows the narrow line ratios taken from table 2 and plotted
against wavelength. The straight line is an unweighted best fit. If
the two narrow line spectra are identical, apart from a linear flux
correction, then the difference between the individual line ratios and
this line reflect the noise in the data.  The standard deviation of
the scatter about the fit is 0.05, shown as dashed lines.

The line ratios for the broad components of H$\alpha$ and H$\beta$ are
also shown and clearly differ from the observed trend in the narrow
lines (filled stars). If the error in the ratio of the broad lines is
similar to that inferred from the narrow lines then we estimate the
significance of the difference to be greater than 3 times the standard
deviation. The ratio of the H$\alpha$ and H$\beta$ narrow components,
which are consistent with the ratios found from the other narrow
lines, have also been included in figure 5 (open circles). Applying
the linear correction found using the narrow line flux ratios to those
for the broad components, we find that
H$\alpha_{B}^{96}$/H$\alpha_{B}^{97} = 0.81 \pm 0.05$ and
H$\beta_{B}^{96}$/H$\beta_{B}^{97} = 0.84 \pm 0.05$, where the error
has been assumed to be equal to the typical scatter of the narrow line
ratios about the fit. However, we expect the errors to be somewhat
larger than this since it is more difficult to measure the flux in
broad lines than in narrow lines. Clearly, we have detected real broad
line variability, but with a considerably smaller factor than the
variations seen in the blue continuum.

Figure 5 also shows the ratio of the fitted continuum observed in July
1996 and January 1997 (see section 4.1) versus wavelength. This shows
that there was a large variation in the blue, but that the variation
in the red is only marginally significant.

\begin{figure*}
\centering
\includegraphics[scale=0.8,bb=30 280 580 670]{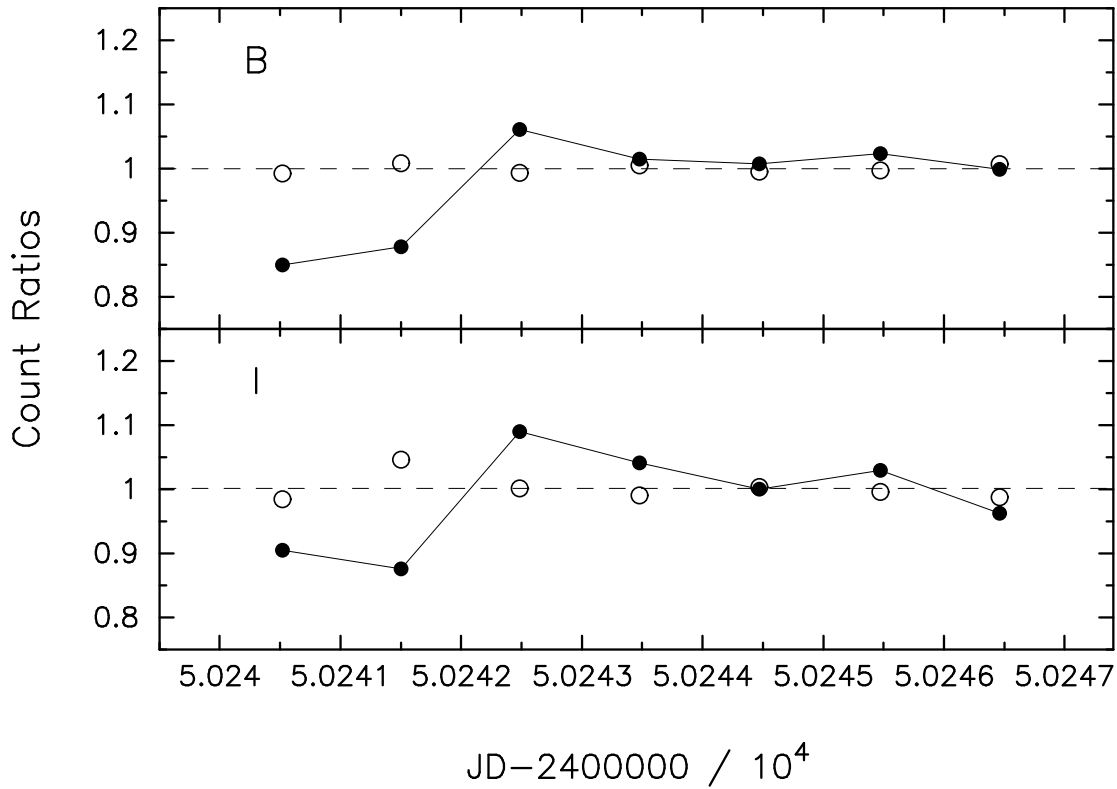}
\caption{JKT I and B relative photometry for the NGC\,4395
nucleus. The count ratios between two field stars (open circles) are
compared with the ratio between the nuclear photometry and one of the
stars (filled circles). The star--star comparison suggests a typical
error of $\sim 2\%$ so the variation in the nuclear photometry is
significant.}
\end{figure*}

Inspecting the spectra by eye, it appears that the broad components of
higher order Balmer lines, such as H$\gamma$ and H$\delta$, have
varied by a larger factor. However, we have not attempted to quantify
this given the noisiness of the data and the problem of blended lines.

An absorption line, identified as CaIIK$\,\lambda3933$, can be seen
near the blue end of the January 1997 spectrum. (CaIIH\,$\lambda3970$
coincides closely with H$\varepsilon$ so is not easily seen). We see
an additional tentative absorption line at $\sim 4055$ \AA, for which
we have no identification.  These lines, as well as some other weak
features, are not seen in the July 1996 spectrum due to the poor S/N
of the data (note that the spectrum in figure 2 has been slightly
smoothed).  The equivalent width of the observed CaIIK line is just
above 1 \AA. The line is quite broad, with a FWHM of $\sim 10$ \AA\,
ruling out the possibility of it being caused by interstellar
absorption. Its profile is slightly asymmetric, which may indicate
more than one component. However, the quality of the data prevents us
from reaching any firm conclusions. There is no evidence of other
important metal lines such as CN\,$\lambda4200$, the G band of
CH\,$\lambda4301$, MgI+MgH\,$\lambda5175$ or NaID\,$\lambda5892$ in
the data. Neither is the 4000 \AA\ break observed, suggesting a very
young stellar population. Bica (1988) shows that in a sequence of
stellar population types from spiral galaxies the equivalent width
(EW) of the CaIIK feature decreases towards younger populations as
well as towards lower metallicities. The absence of the 4000 \AA\
break means that of Bica's templates only groups S6 and S7 can
apply. For S7 most of the light is thought to come from populations of
age $\sim 10^{8}$ years, and still has a EW of CaIIK $\sim 3.5$
\AA. Although we cannot separate the effects of age and dilution, it
seems likely that the population is young (conservatively $< 1$ Gyr)
and has a dilution of at least 70\%.  

Our detection of the CaIIK absorption line suggests that other stellar
features characteristic of young clusters may also be detectable.
Among these, the near infrared CaII\,$\lambda8498,8542,8662$ triplet
should be the strongest (Terlevich et al.\ 1990).  A detection of the
IR CaII triplet would provide confirmation of the presence of the
nuclear young cluster plus direct measurement of its velocity
dispersion thus its dynamical mass plus an estimate of dilution factor
in the near infrared.

\subsection{Broad-band variability}

After careful examination of the JKT $B$ and $I$ images, no evidence
was found for nuclear variability within any of the two hour observing
windows each night.  Therefore, the data for each night were
averaged. To minimise any colour-dependent effects due to differences
in the intrinsic spectral energy distribution of the nucleus, galaxy
and stars, only data from frames for which the airmass was less than
1.5 were used to construct the nightly averages. The average airmass
is very similar for each night as the observing period was similar in
UT.

The average $B$ and $I$ band CCD count ratios for two stars and for
the nucleus and one of the stars are shown in figure 6. The values
have been normalised to unity using the data from the last four nights
of the run. Based on the star ratios, we conservatively estimate the
one-sigma uncertainty for flux variability to be 2 per cent. The
nucleus is clearly variable over the first few nights, with the
largest change being a brightening by about 20 percent from night two
to three. The changes are similar in form and amplitude in both
bands. Other small-amplitude variations are possibly present during
the second half of the week, although these are not of high
significance.

Absolute photometric magnitudes were derived using several flux
standards and field stars all observed at very low airmass ($< 1.1$)
so colour differences are insignificant.  This implies nuclear (galaxy
subtracted) magnitudes of $B = 16.8$ and $I = 15.8$ for NGC\,4395
during the JKT run. We estimate 1-sigma errors of $\pm 0.1$
magnitudes.  This result is fairly consistent with the trend shown by
our spectroscopy: the continuum becomes harder when brighter.
Filippenko \& Sargent (1989) quote a B magnitude of $\sim 17.3$ (0.42
mJy) based on the flux density at 4400 \AA\ from their spectroscopic
data.  From our spectroscopy shown in figure 2, the fluxes at 4400
\AA\ imply B magnitudes of 18.0 and 17.5 for July 1996 and January
1997 respectively. This implies a historical flux variation by a
factor $\sim 2$, and strongly supports our detection of variability in
NGC\,4395.

\subsection{Analysis of {\it HST\/} images}

\begin{figure*}
\centering
\includegraphics[scale=0.9,bb=100 140 520 700]{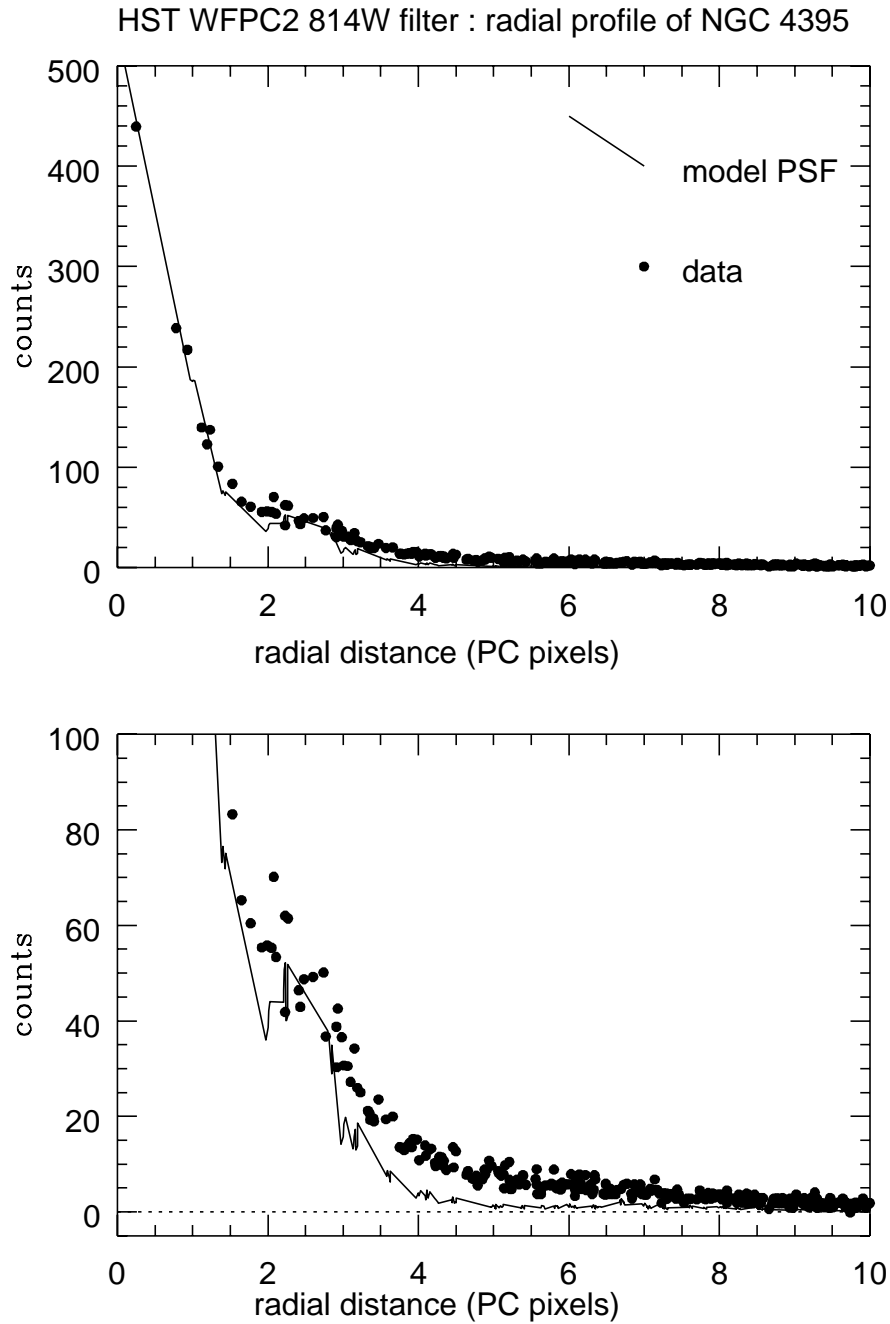}
\caption{Comparison of the radial profile of the NGC\,4395 nuclear
emission and model TinyTim psfs.  The nucleus of the galaxy was imaged
onto the {\it HST\/} PC 2 using the F814W filter. The bottom panel
shows a detail of the top panel, so that the extended component can be
seen more clearly. Assuming a distance to NGC\,4395 of 5.21 Mpc, the
plate scale is 0.046 arcsec per pixel (1.16 pc per pixel).}
\end{figure*}

Photometry in standard bands was derived from the {\it HST\/}
observations using a circular aperture of 1 arcsec diameter. The
background emission was negligible. This gave values of $B=16.91$ and
$I=16.23$, comparable to the values from the JKT broad-band imaging,
further supporting evidence of variability.

{\it HST\/} imaging has been previously discussed by Filippenko, Ho \&
Sargent (1993).  However those observations were taken by WFPC1,
whereas the data we discuss here ware taken with WFPC2.  We therefore
examine the images for any sign of resolved structure.  Model point
spread functions (PSFs) for both the F450W and F814W filters at the
relevant chip positions were calculated using Tiny Tim
software. Standard {\sc iraf} routines were then used to find the
centroids and radial profiles of both the real and model data.
Comparison of {\it HST\/} data with model PSFs in the very central
regions is very sensitive to centroid location with respect to pixel
centres, as even the PC is somewhat undersampled. We do not attempt a
proper model test here, but rather have simply scaled the model PSF by
eye to match the real data at radii of 1-2 pixels, in order to look
for evidence of extended structure.

The F814W data have a clear peak pixel, so a direct comparison with a
PSF is relatively secure.  The comparison with the PSF is shown in
figure 7. It can be seen that the core of the PSF matches the data
very well, so we confirm the finding of Filippenko, Ho \& Sargent
(1993) that most of the flux from the nucleus of NGC\,4395 comes from
an unresolved point source. We have not performed a fit, but an
intrinsic FWHM of more than about half a pixel would have been easily
detected. At our adopted distance of 5.21 Mpc this corresponds to a
physical size of 0.6 pc.

In the wings of the profile, however, the source has a clear excess
above the PSF. Although this is a very small number of counts, it is a
large factor above the predicted wings, so it is unlikely to be
accounted for by a slightly different normalisation of the PSF without
disagreeing badly with the core. In this region the Tiny Tim PSF
should be very reliable and the spacecraft jitter during these
observations was only 0.14 pixels.  There is, then, evidence of
diffuse emission surrounding the core of NGC\,4395. Without proper
modelling, which we defer to a later paper, it would be unwise to
quantify the excess, but very roughly the integrated diffuse flux
could be as much as 10 per cent of the core flux, and its physical
scale is a diameter of $\sim$ 18 pixels = 0.8 arcsec = 20 pc, similar
to the size of the extended [OIII] emission detected by Filippenko, Ho
\& Sargent (1993) - see also Matthews et al.\ (1996). Indeed it is
quite possible that at the 5-10 per cent fraction of the core that we
are considering here, the diffuse light in the F814W filter is
entirely due to nebular emission. On the other hand, the detection of
an underlying young stellar component (see section 3.2) could explain
the diffuse emission as contamination by a nuclear stellar cluster.

In the blue (F450W) observation, the data has two bright pixels and
any point source is probably centred somewhere between these. The type
of crude analysis we used above is, therefore, even less reliable, and
so we defer detailed discussion of this image. Very roughly however,
the data are at least consistent with the same story - an unresolved
core and diffuse emission at the 10 per cent level.

\section{Discussion}

\begin{figure*}
\centering
\includegraphics[scale=0.9,bb=30 360 580 700]{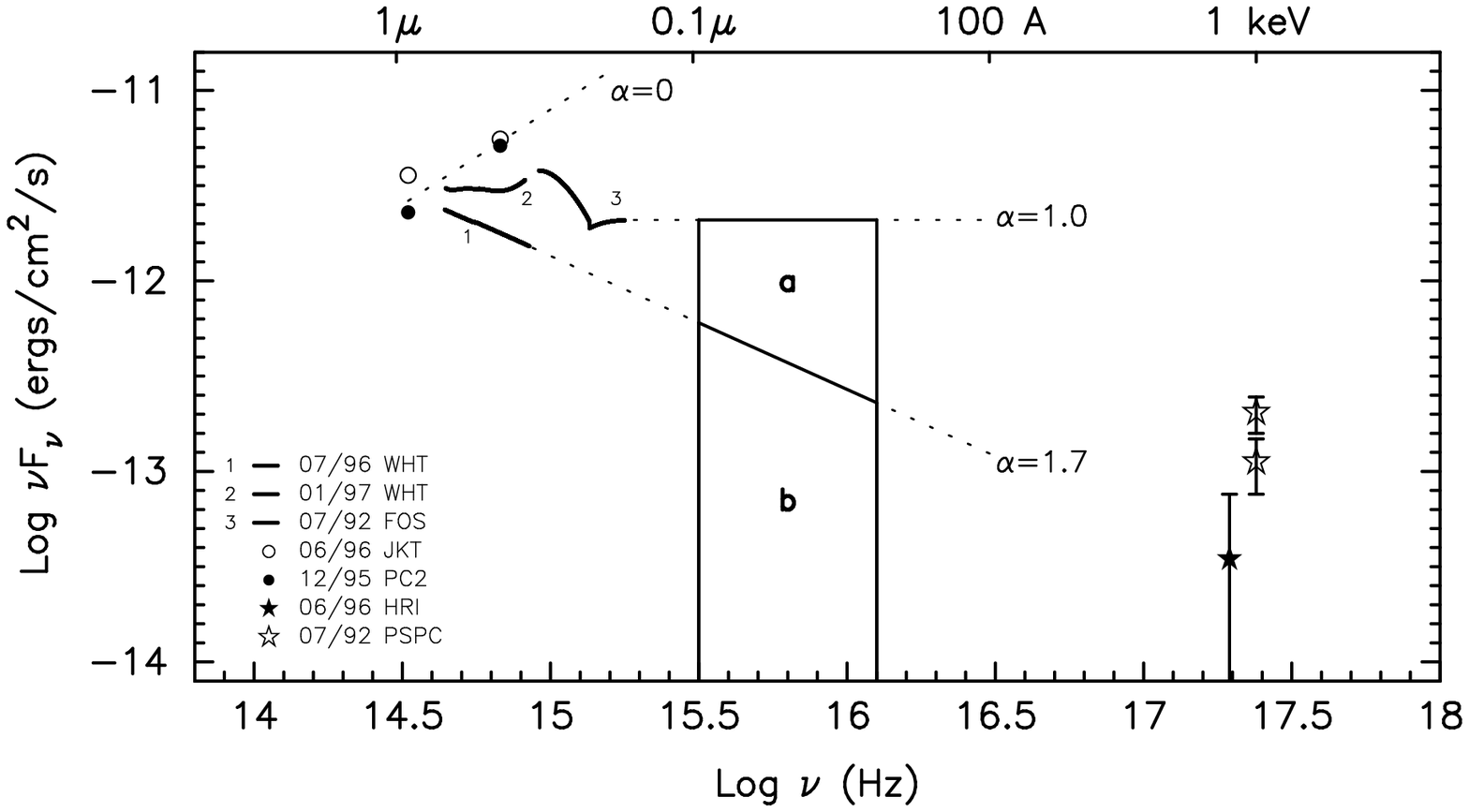}
\caption{Spectral energy distribution of the nucleus of NGC\,4395.
Dotted lines have been drawn as examples of power law functions with
energy spectral indices $\alpha = 0, 1$ and 1.7. The ionizing
continuum (between $\nu = 3.3 \times 10^{15}$ and $\nu = 10^{16}$ Hz)
has been represented by the areas a and b (see text). The optical
spectra (lines 1 and 2) were obtained in July 1996 and January 1997
and show a low and high state of activity.  The FOS UV spectrum (line
3) was retrieved from the WHT public archive (it has been published by
Filippenko, Ho \& Sargent (1993)). B and I broad band observations are
shown as open circles (JKT data) and filled circles ({\it HST\/}
data).  The best estimate of the measured {\it ROSAT\/} HRI X-ray flux
is indicated with an filled star. Archive {\it ROSAT\/} PSPC fluxes
are indicated with open stars. All the X-ray data are given with
$2\sigma$ error bars.  The figure key shows the dates (month/year) of
the different observations. Note from table 1 that the X-ray
observations are not simultaneous with any other data.}
\end{figure*}

\subsection{Spectral energy distribution}

For the determination of the Spectral Energy Distribution (SED) of
NGC\,4395, ultraviolet {\it HST\/} spectra obtained with the Faint
Object Spectrograph (FOS) in July 1992 were retrieved from the public
archive. The original data were published by Filippenko, Ho \& Sargent
(1993). 

To determine the featureless UV and optical continuum a low degree
polynomial was fitted to the spectra, allowing for the rejection of
all the emission-line features during the fitting procedure. The
resulting UV continuum will include broad quasi-continuum such as
Balmer continuum and FeII emission-lines. Some starlight contamination
is expected as stellar signatures have been observed in the nuclear
emission but it is only of the order of 10\% of the total flux at 4000
\AA\ (see sections 3.2 and 3.4).

For the X-ray data an effective energy for the {\it ROSAT\/} HRI
bandpass of 0.8 keV was deduced assuming a power law spectrum of index
$\alpha = 1.5$. For the PSPC the value adopted was 1.0 keV.

Figure 8 shows the observed SED for NGC\,4395. Dashed lines have been
drawn as examples of power law functions with spectral indices $\alpha
=$ 0, 1 and 1.7 ($f_{\nu} \propto \nu^{-\alpha}$). Several striking
features are evident in the SED plot. The change in the continuum
shape between our two spectroscopic observations is dramatic. The
featureless optical continuum obtained in July 1996 is very steep,
with a best fitted spectral index $\alpha \approx 1.7$. The
extrapolation of this continuum to shorter wavelengths seems to agree
with the observed X-ray luminosities. Compared with more luminous AGN,
NGC\,4395 seems to be a very quiet X-ray source, unless the PSPC and
the HRI observations were made during extremely low activity states.

The optical continuum looks flatter and brighter by January 1997 with
a best fitted $\alpha \approx 1.1$, but starts to raise for
wavelengths shorter than 4000 \AA. The {\it HST\/} UV data from 1992
shows a sharp break at $\sim$ 2200\AA. After the break the UV spectrum
seems to agree with an $\alpha = 1$ spectral distribution. The bump
observed at $\sim$ 3000 \AA\ nicely matches the optical observations
of 1997. However, due to the uncertainties in the spectral absolute
flux calibration and the 4 year gap between the optical and the UV
observations, the match could be just fortuitous.

The 3000 \AA\ UV bump in figure 8 resembles the `Small Blue Bump'
observed in many other classic AGN (Wills, Netzer \& Wills 1985),
which is believed to be a blend of broad permitted FeII lines and
Balmer continuum (Bac). The UV excess corresponds to $\sim 4.0 \times
10^{-13}$ ergs cm$^{-2}$ s$^{-1}$ (obtained by integrating below the
SED beteween 2220 to 4500 \AA\ and subtracting the flux of the
underlying continuum approximated as a power law), giving an
estimation for the UV excess/H$\alpha$ ratio of $\sim$ 2.5. Observed
ratios for more powerful AGN range from $\sim$ 2 to 5, in agreement
with our observations (Edelson \& Malkan 1986; Malkan 1983). These
values are much larger than the model predictions for Case B
recombination (Kwan \& Krolik 1981), suggesting that perhaps more than
the 50 per cent of this UV excess comes from FeII emission. The lack
of observed FeII features in the optical range in NGC\,4395 is not
unusual. Wills, Netzer \& Wills (1985) and Netzer et al.\ (1985) have
shown that strong UV FeII emitters can show extremely weak optical
FeII lines.

The $\sim$ 2200 \AA\ break can also be explained as substantial
internal reddening in NGC\,4395. Using the extinction curve by
Cardelli, Clayton \& Mathis (1989) we find that a visual absorption
$A_{V} \la 0.4$ is required to produce the dip at 2200 \AA. The dust
responsible for this extinction is probably not located in the Broad
Line Region (BLR) since it is unlikely to survive the high densities
and temperatures. Nor does the observed Balmer decrement support the
presence of significant dust in the NLR: from table 2 we find
H$\alpha_{N}$/H$\beta_{N} \sim 2.5$, while an extinction $A_{V} \sim
0.4$ implies H$\alpha_{N}$/H$\beta_{N} \sim 3.3$. Rowan-Robinson
(1995) claims that the dust responsible for optical and UV reddening
in quasars, with $A_{V} \sim 0.1 - 0.5$, is diffuse material located
in the NLR or associated with the interstellar medium of the parent
galaxy. The later seems to agree with our observations, although an
$A_{V} \sim 0.4$ is a slightly high amount of extinction for a nearly
face on galaxy.

We seem to have caught NGC\,4395 in three different states. The broad
band JKT data and the {\it HST\/} images show a bright and blue source
with $\alpha \sim 0$, similar to luminous quasars. The January 1997
and 1988 data show a state with $\alpha \sim 1$ and a much less
obvious `Big Blue Bump'. Finally, in July 1996, NGC\,4395 shows a
steep spectrum with $\alpha \sim 2$. The relative strength and slope
of the Blue Bump can vary widely amongst quasars and Seyfert galaxies
(Elvis et al.\ 1994; Walter \& Fink 1993; Puchnariewicz et al.\
1996). In the NGC\,4395 medium state, the observed continuum with
$\alpha \sim 1$ is in fact quite similar to that seen in many Seyfert
galaxies (e.g., Edelson and Malkan 1986; Kriss et al.\ 1991). The low
state spectrum of July 1996 is much more unnusual, but not
unprecedented - for example, the ultraluminous IRAS galaxy F10214+4724
and the archetypal Type 2 Seyfert NGC\,1068 both show steep spectra
throughout the near-IR to UV (see Lawrence et al.\ 1994 and references
therein). More recently two other extremely low luminosity broad-line
AGN have been found to have a steep optical-UV spectrum - M81 (Ho,
Filippenko \& Sargent 1996), and NGC\,4579 (Barth et al.\ 1996). In
both these cases the X-ray luminosity is very strong, with
$\alpha_{ox}$ = 0.92 and 0.86 respectively.  However, for NGC\,4395,
the continuum seems to continue falling steeply all the way to soft
X-rays ($\alpha_{ox} \sim 1.9$ -- see section 4.6).

\subsection{Optical spectroscopic variability}

\begin{figure*}
\centering
\includegraphics[scale=0.8,bb=60 420 560 700]{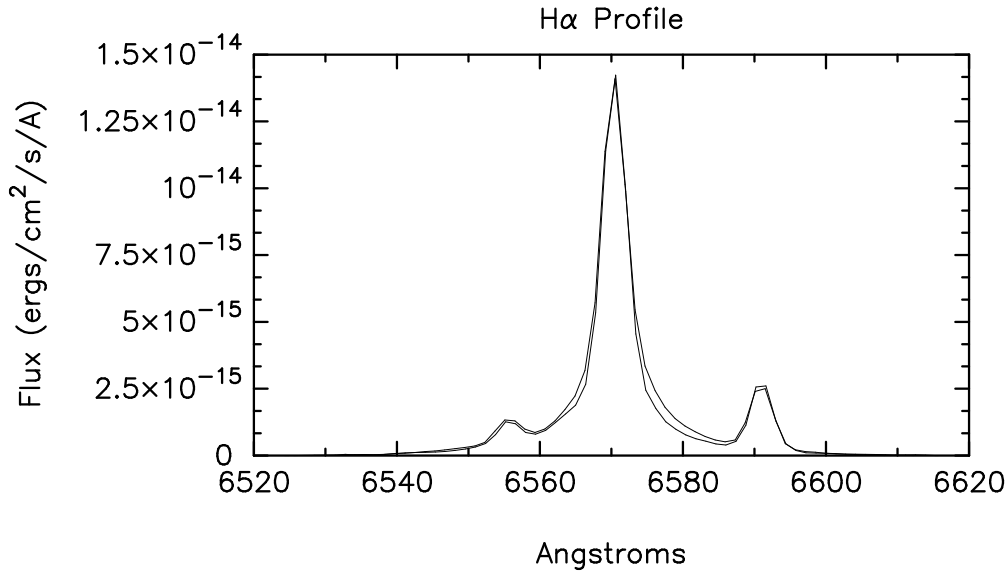}
\caption{H$\alpha$ + N\,II line profiles for the high and low state of
NGC\,4395. The continuum have been subtracted from both spectra and a
scale factor of 1.2 has been applied to the low state data.}
\end{figure*}

For the first time optical variability has been reported for
NGC\,4395. Previous observations had given negative results for any
change in the continuum level or line fluxes (Shields \& Filippenko
1992), although this was not quantified. Our data show a substantial
change in the optical continuum (3700 - 6700 \AA), as can be seen in
figure 2. It seems that NGC\,4395 moves between high and low activity
states, characterized by spectral distributions with indices $\alpha
\approx 1.0$ and $\alpha \approx 1.7$, respectively. The broad-band
data suggest an even higher state with $\alpha \sim 0$.

As was mentioned earlier, a significant variation in the H$\alpha$ and
H$\beta$ broad line fluxes has been detected, but with an amplitude
much smaller than the variation exhibited by the continuum.
Variability studies for local Seyfert galaxies show that the amplitude
of the flux variations changes from line to line, with the high
ionization lines showing the largest variations (Carone et al.\ 1996;
Clavel et al.\ 1991; Kassebaum et al.\ 1997). While a variation of
$\sim$ 1.5 is normal for the total H$\beta$ flux, a factor of up to 16
has been observed for HeII$\lambda4686$. From table 3 we see that the
(not corrected - see section 3.2) ratio of the 1996 fluxes to 1997
fluxes is 1.2 for H$\beta$ and 1.3 for H$\alpha$.

Rosenblatt et al.\ (1992) carried out optical spectroscopic monitoring
of 13 Seyfert galaxies with $5 \times 10^{39} <$ L(H$\beta$) $< 4
\times 10^{42}$ ergs s$^{-1}$ from 1979 to 1984. They found strong
evidence that variability in the continuum and H$\beta$ occur on time
scales of 90 days or less. They also found that less luminous galaxies
are more strongly variable in their lines and continuum.  Our report
of optical variability for NGC\,4395 shows that extremely low
luminosity Seyfert galaxies do vary, but it is still unclear whether
they follow the pattern shown by the more luminous objects. Rosenblatt
et al.\ also show that in 8 galaxies peak-to-peak changes in H$\beta$
fluxes were 100 per cent -- 200 per cent, while the continuum varied
by 200 per cent -- 350 per cent for 12 of the 13 galaxies, in
agreement with our observations for NGC\,4395.

No change in the line profiles is observed between the two epochs, as
can be seen in figure 9, where the H$\alpha$ + N\,II lines from July
1996 and January 1997 have been plotted together, after continuum
subtraction and a scaling of the former by a factor of 1.2. It has
been shown for NGC\,5548 that while emission-line fluxes vary with
changes in the ionizing continuum, the line profiles do not reflect
these variations (Wanders \& Peterson 1996; Kassebaum et al.\
1997). Instead, we expect the time-scale for line profile changes to
correlate with the BLR kinematical evolution times-cale.

As has been noted, it is a common property of Seyfert galaxy
variability that the optical continuum becomes harder when the source
becomes brighter. However, it has always been worrying that this might
be due to stellar contamination in the red. For NGC\,4395, the
surrounding bulge is of very low surface brightness and Filippenko, Ho
\& Sargent (1993) stress the featurless nature of the continuum,
suggesting that any contribution from an old stellar population is
small. Our discovery of CaIIK absorption (section 3.2) and a diffuse
component to the nuclear broad-band emission (section 3.4) suggests
that a small but significant fraction of the light (10--20\%) comes
from a very young stellar cluster, but this is unlikely to produce the
colour dependence observed in variability. An important test will be
high S/N spectroscopy in the low state, when the CaIIK absoprtion
should be larger.

\subsection{Fast variability}

In the X-rays, we have seen a factor two variation in 15 days. By the
standards of local AGN, this is not particularly fast (for example,
NGC\,4051 can change by a factor of two in half an hour - see
Papadakis \& Lawrence 1995). Likewise, optical spectroscopic
variability by a factor of two in six months is quite reasonable for
local Seyfert galaxies like NGC\,5548. Luminous quasars variations are
slower, taking typically several years to vary by a factor of two -
see for example the light-curves in Hawkins (1996).  However we have
evidence in two ways that NGC\,4395 varies in a more dramatic fashion
than classical Seyfert galaxies.

First, we note that the last night of our broad-band imaging (JKT)
observations were taken only 32 days before the night in which we
found the spectroscopic low state. During this time the B-band flux
decreased by a factor three. Variations by a factor of three in the UV
over 50 days certainly occur in Seyfert galaxies - for example in the
well known 1989 monitoring of NGC\,5548 (Clavel et al.\ 1991). Over
the same period, the blue-optical flux (at 4870 \AA) followed the
same pattern of variation, but with a much smaller amplitude, about a
factor 1.5 (Peterson et al.\ 1991).  The largest optical variations
seen in the four year 1988 - 1992 optical monitoring of NGC\,5548
(Peterson et al.\ 1994) are a factor two over about 100 days.

The second piece of evidence that NGC\,4395 varies more dramatically
and/or more quickly is that we have seen a 20 per cent change from one
night to the next during our broad-band imaging observations. Very few
short-times-cale monitoring campaigns have been carried out. Optical
monitoring of the low luminosity Seyfert NGC\,4051 by Done et al.\
(1990) found the $B$-band flux to be constant within 1 per cent over a
whole week. Likewise, the two week long monitoring of NGC\,4151 by
Edelson et al. (1996) showed that variations at 5125 \AA\ were 1 per
cent or less. During the same period however, UV variations were
considerably larger, with a normalised variability amplitude of 9 per
cent at 1275 \AA, and one particular event showing a rise of $\sim 20$
per cent over 1 day.

There is then tentative evidence that either NGC\,4395 varies
characteristically more quickly or with larger amplitude than more
luminous AGN.  However for the intra-week monitoring in particular we
may have seen a freak event. Repeated monitoring is clearly needed to
see what may be typical. UV and fast X-ray monitoring would also be of
great interest.

\subsection{The ionizing continuum}

As noted in section 4.1, the blue bump in NGC\,4395 is occasionally
strong, and sometimes very weak.  In this section we ask whether the
implied ionising continuum is luminous enough to produce the observed
broad Balmer lines. This depends sensitively on how one extrapolates
into the UV. If we assume a power-law spectrum $L_{\nu} =
L_{\nu_{\circ}} (\nu / \nu_{\circ})^{-\alpha}$, where
$L_{\nu_{\circ}}$ is the monochromatic luminosity per unit frequency
at the hydrogen ionisation edge $\nu_{\circ}$, then extrapolating to
infinity gives the number of ionising photons $N_{ion} =
L_{\nu_{\circ}} / h\alpha$. (Some authors extrapolate to the He edge
at 228 \AA\ but this makes relatively little
difference). $L_{\nu_{\circ}}$ can in turn be estimated by
extrapolating from the observed optical/UV continuum. Then on standard
Case B assumptions, and with a unity covering factor and assuming that
all the ionising photons are absorbed $N_{ion}/N_{H\alpha} =
2.2$. More realistic conditions should change this predicted value by
less than a factor two.  Relaxing the other assumptions (such as
complete coverage) makes $N_{ion}/N_{H\alpha}$ larger.

For the medium state, we can use the H$\alpha$ luminosity observed in
January 1997, and calculate $N_{ion}$ by extrapolating the {\it HST\/}
UV luminosity at 2000 \AA\ with $\alpha = 1$. (This corresponds to the
sum of areas a+b in figure 8 where, for the purposes of the plot, we
have adopted the He edge as the upper limit for the ionizing
continuum; the actual determination of $N_{ion}$ was done integrating
to infinity). This gives $N_{ion}/N_{H\alpha} = 1.5$ which, within the
uncertainties of this calculation is just consistent with H$\alpha$
being produced by photo-ionisation.  For the low state, we can use the
H$\alpha$ luminosity observed in July 1996 (which is only 0.8 times
smaller than the H$\alpha$ luminosity in the high state). If we
optimistically extrapolate from the blue end of our optical spectrum
with $\alpha = 1$ the result is very similar to the case above.
However, if the observed steep spectrum continues falling as
$\alpha=1.7$ (area a in figure 8) we find $N_{ion}/N_{H\alpha} = 0.4$,
and conclude that the deduced ionising luminosity fails to explain the
observed broad H$\alpha$ by a substantial factor.

Such a deficit of ionising photons has been claimed for Seyfert 2
galaxies (e.g., Wilson, Ward \& Haniff 1988; Kinney et al.\ 1991) and
for the extended emission-line regions of radio galaxies (e.g.,
Robinson et al.\ 1987), where it has been used to argue for an
obscured and/or anisotropic continuum. To our knowledge such a deficit
has rarely if ever been claimed for traditional luminous broad-line
objects, i.e., Seyfert 1s and quasars - indeed the excess of available
continuum to lines is often used to deduce that the BLR covering
factor is much less than 1 (Yee 1980, Shuder 1981). One exception we
are aware of is that Filippenko (1985) argued that extrapolating the
optical continuum of the broad line radio galaxy Pictor\,A yields a
deficit of a factor three.  More recently, Barth et al.\ (1996)
perform a similar calculation for the very low luminosity broad line
object NGC\,4579, which also has a steep UV continuum.  They find that
the extrapolated ionising continuum is marginally sufficient to
explain the observed narrow lines, but fails to explain broad
H$\alpha$ by a substantial factor.

Obviously if the ionising continuum cannot explain the observed broad
emission-lines, this is a potentially important result, and may imply
that the continuum is anisotropic, or that an extra heating source is
required for Balmer lines in AGN, such as mechanical heating of some
kind. However, given that H$\alpha$ varies, that the UV very likely
also varies, and that we do not have simultaneous optical and UV data,
we cannot yet make such a bold statement. Further monitoring,
especially in the UV, is very important.

\subsection{Black hole mass in NGC\,4395}

We can crudely estimate the bolometric luminosity of NGC\,4395 by
assuming a simplified shape for the whole spectral energy
distribution. Long wavelength observations in principle help constrain
the SED below 1 $\mu$m. Radio VLA observations of a compact source
coincident with the nucleus of NGC\,4395 give a flux of $1.24 \pm
0.07$ mJy at 20 cm and $0.56 \pm 0.12$ mJy at 6 cm. IRAS observations
at 12 $\mu$m, 25$\mu$m and 60$\mu$m show cirrus-like emission, most
probably coming from cold dust heated by the interstellar radiation
field within the NGC\,4395 disk, rather than warmer dust heated by the
nuclear source.

Without far-infrared observations of the nucleus to constrain the
bolometric luminosity, we will assume two very simple models: (1) a
power law with index $\alpha$ = 1.7, normalised at 6800 \AA\ to match
our optical spectroscopy, and with cut-offs at 20 $\mu$m and 2 keV;
(2) the same as (1) but assuming $\alpha$ = 1. The bolometric
luminosities are found to be $L_{Bol}^{(1)} = 1.18 \times 10^{41}$
ergs s$^{-1}$ for the first case and $L_{Bol}^{(2)} = 1.21 \times
10^{41}$ ergs s$^{-1}$ for the second. The two values are fortuitously
similar, hiding the fact that in the first case nearly all the
luminosity is in the mid-IR, and in the second case in the UV. If we
had chosen 100 $\mu$m as the cut off in the far infra-red, the
bolometric luminosity increases by almost an order of magnitude for
case (1), while it remains almost unchanged for case (2).  Note also
that if the infrared emission is largely reprocessed energy, then for
some source geometries including both the raw and reprocessed
components may overestimate the nuclear luminosity . We will adopt
$L_{Bol} = 1.2 \times 10^{41}$ ergs s$^{-1}$ as a representative value
for the bolometric luminosity.

If $v$ is the velocity dispersion of virialized clouds within a
distance $R$ of the nucleus then the central mass of an AGN is $M =
Rv^{2}$/G. While $v$ can be estimated from the half width zero
intensity of the broad line components ($v \sim \sqrt{3}/2$
FWZI(H$\beta$), Wandel 1991), the value of $R$ is less certain. In the
last 10 years reverberation mapping has made it possible to estimate
the size of the BLR in a direct way, in contrast with previous
estimations through AGN standard photoionization models which had
overestimated $R$ by about an order of magnitude (Peterson 1994; Maoz
1994).

Reverberation mapping also appears to confirm a luminosity-size-mass
relationship consistent with $R \propto L^{1/2}$ (Kaspi et al.\
1996b). Such a relationship is theoretically expected given two
somewhat na\"{\i}ve assumptions: (1) the shape of the ionising
continuum is the same for all AGN, and (2) the BLR is characterized by
the same value in all cases for the product of the ionising parameter
U (the ratio of the density of ionized atoms to ionizing photons) and
the cloud density $n_{e}$. The best fit slope from AGN data covering
two orders of magnitude in luminosity gives $R = 1.4 \times 10^{-24}
L_{0.1-1\mu m}^{1/2}$ pc, where $L_{0.1-1\mu m}$ is the 0.1 -- 1
$\mu$m luminosity in units of ergs s$^{-1}$ (Netzer \& Peterson
1997). However, the emission spectrum of LINERs, which are regarded as
low luminosity versions of the more powerful AGN, have been explained
with a value for the ionization parameter 10 times smaller than that
inferred for Seyfert galaxies (Ferland \& Netzer 1983). Even if all
Seyfert 1 galaxies (including NGC\,4395) can be characterised by a
particular photoionization regime, any extrapolation of the physical
conditions derived for the more luminous objects to very low
luminosity sources may introduce major errors in our calculations.

Bearing in mind all the limitations discussed above we find, from our
spectra of NGC\,4395, that $L_{0.1-1\mu m} \sim 1.7 \times 10^{40}$
ergs s$^{-1}$, implying $R = 5.7 \times 10^{14}$ cm. For FWZI(H$\beta)
\sim 4100$ km s$^{-1}$ the mass of the central object in NGC\,4395 is
$\sim 1.1 \times 10^{36}$ kg, or $5.4 \times 10^{5}$ $M_{\sun}$. The
corresponding Eddington luminosity is $L_{Edd} = 7.0 \times 10^{43}$
ergs s$^{-1}$. From the estimates of $M$ and the bolometric luminosity
$L_{Bol}$ we find that the central source is emitting at $\sim 1.7
\times 10^{-3} L_{Edd}$.

Otherwise, if the distance to the BLR does not scale as $L^{1/2}$ to
very low luminosities but, instead, approaches a minimum value
($R_{min}$), then the central mass in NGC\,4395 could be much
larger. The smallest observed values of $R$ correspond to 0--3
light-days for NGC\,4151 (Kaspi et al.\ 1996a) and 4 light-days for
NGC\,4593 (Dietrich et al.\ 1994). A $R_{min} \sim 1$
light-day would imply a black hole mass of $2.5 \times 10^{6}$
$M_{\sun}$ and NGC\,4395 would be extremely underluminous, emitting at
just $\sim 4 \times 10^{-4} L_{Edd}$.

\subsection{X-ray loudness}

To compare the X-ray flux with continuum emission at other wavelengths
we will use the spectral indices $\alpha_{ox}$ and $\alpha_{ix}$,
where the index $\alpha_{ox}$ is defined between 2500 \AA\ ($\log\nu =
15.08$) and 2keV ($\log\nu = 17.68$), and $\alpha_{ix}$ between 1
$\mu$m ($\log\nu = 14.48$) and 2 keV.  Using the observed flux at 2500
\AA, and extrapolating the optical spectroscopy to obtain $F_{1\mu m}$
we find that using the range of observed X-ray fluxes gives
$\alpha_{ox} = 2.03 - 1.76$ and $\alpha_{ix} = 1.85 - 1.63$, where the
monochromatic X-ray fluxes at 2 keV have been determined from the
broad-band observations assuming a power law index of 1.5 between 0.1
and 2.4 keV.

Values of $\alpha_{ox}$ for AGN (Mushotzky \& Wandel 1989; Laor et
al.\ 1994; Laor et al.\ 1997 ; Walter \& Fink 1993; Barth et al.\
1996; Mushotzky 1993) are normally regarded as a good estimate of the
size of the `Blue Bump' relative to the X-ray emission.  Typically,
for quasars and Seyfert 1 galaxies, $\alpha_{ox} \sim 1.5$. For LINERs
a much smaller value is found ($\alpha_{ox} \sim 0.9$) due to the
lower UV fluxes relative to X-rays. NGC\,4395 has a steep
$\alpha_{ox}$ which is the result of the low X-ray flux rather than
powerful UV emission. A similar result is found amongst other low
luminosity Seyfert galaxies (Koratkar et al.\ 1995) which are faint in
the UV, but even less bright in the X-rays, giving $\alpha_{ox} \sim
1.6$.

The power law index $\alpha_{ix}$ better characterizes X-ray loudness,
since the flux at 1 $\mu$m is less affected by either internal
absorption or the intrinsic magnitude of the UV power.  The value of
$\alpha_{ix}$ found for NGC\,4395 ($\sim 1.7$) is much larger than
that found for more luminous radio-quiet AGN ($\sim 1.3$, Lawrence et
al.\ 1997; see Mushotzky \& Wandel 1989 for a power law slope of 1.27
between 7500 \AA\ and 2 keV). This shows that the overall spectral
shape for NGC\,4395 drops rapidly from the optical to the soft X-rays.

Assuming a mean $\alpha_{ix}= 1.74$ we find that NGC\,4395 is about 25
times less X-ray loud than more powerful radio-quiet AGN.  Adopting an
absorption cross section per hydrogen atom at 1keV of $\sim 2 \times
10^{-22}$ cm$^{2}$ (Morrison \& McCammon 1983) we find that a hydrogen
column density of $\sim 1.4 \times 10^{22}$ cm$^{-2}$ is required to
explain the X-ray deficit. A normal interstellar Galactic gas-to-dust
ratio N(HI)/E$(B-V) = 5 \times 10^{21}$ atoms cm$^{-2}$ mag$^{-1}$
(Sauvage \& Vigroux, 1991), and a Galactic extinction law give A$_{V}
\sim 8.4$ mag, which is not supported by the ratio of the narrow
components of the Balmer lines (see table 2) that show a Balmer
decrement of H$\alpha_{N}$/H$\beta_{N} \sim$ 2.5. However, it is quite
usual that local Seyferts with large X-ray columns do not have the
corresponding line reddening, suggesting absoprtion by dust free gas.
Examples of X-ray quiet quasars can be found in Laor et al.\ 1997. It
is unknown if these objects are intrinsically faint, or absorbed but
otherwise normal quasars. However, the observed Balmer decrement of
the broad component, $<$H$\alpha_{B}$/H$\beta_{B}> \sim 6$, does imply
an $A_{V} \ga 1$ if an extinction free ratio of 4.5 is adopted as
typical for AGN (Osterbrock 1989). This, together with the lack of
ionizing flux noted in the previous section and the break of the UV
continuum at $\sim$ 2200 \AA\ could imply that some amount of
absorption and extinction are indeed occurring somewhere in the line
of sight towards the central source.

\subsection{Comparison with other low luminosity AGN}

The LINER nucleus in M\,81 is a very well studied low luminosity
AGN. Compact X-ray and radio sources coincident with the nucleus have
been detected (Fabbiano 1988, Bartel et al.\ 1982). It has an
H$\alpha$ broad component with a flux 20 times fainter than the flux
detected in NGC\,4051, the dimmest `classical' Seyfert galaxy
(Filippenko \& Sargent 1988). Compared to M\,81, the
luminosity of the mean H$\alpha$ broad component for NGC\,4395 taken
from table 3 is $\la 10$ times smaller. Another example of a LINER
nucleus that shows AGN-like properties (broad H$\alpha$ and X-ray and
UV variability) is NGC\,4579 (Barth et al.\ 1996)

Estimates for M\,81 show that if photoionization is responsible for
its LINER activity, values of U $= 10^{-2.8}$ and $n = 10^{9}$
cm$^{-3}$ are required, and a radius for the BLR of $1.2 \times
10^{17}$ cm can be inferred (Ho, Filippenko \& Sargent 1996). For
NGC\,4395 the size of the BLR found in section 4.5 is $\sim 200$ times
smaller (under the assumption that U and $n$ are as inferred from more
powerful AGN). The central source in M\,81 is emitting at $\sim 0.06
\pm$ 0.04 per cent of the Eddington limit (Ho, Filippenko \& Sargent
1996), a factor 3 smaller than the value found for NGC\,4395. As has
been suggested before, the low ionization lines observed in LINERs
could be caused by small values of U and the consequent reduction of
the ionizing photon flux, probably as a result of a larger distance
between the central source and the emitting clouds (Halpern \& Steiner
1983; Ferland \& Netzer 1983). 

The shape of the optical featureless continuum during the low state of
NGC\,4395 resembles the steep UV continuum observed in M\,81 (Ho,
Filippenko \& Sargent 1996) for which a spectral index close to 2 was
derived. Unfortunately, the shape of the continuum of M\,81 in the
optical range is not well known because of the heavy contamination by
starlight. Maoz et al.\ (1998) have recently studied the UV spectra of
seven LINERs. They found that two of the nuclei (M\,81 and NGC\,4579)
show strong broad emission lines. However, at least in other three
cases the UV continuum is dominated by stellar features from massive
young stars. They suggest that LINERs could be formed by two
distinctive populations: a first group where young massive stars are
responsible for the photoionization of the line emitting gas, and a
second group where the central source is a genuine AGN. The possible
UV variability detected in NGC\,4579 supports this view (Barth et al.\
1996).

Optical variability has not been reported for M\,81 (Ho, Filippenko \&
Sargent 1996). The H$\alpha$ light-curve shows only slight variation
(10 per cent -- 15 per cent), the reality of which is unclear, given
the complexity of the observations. There is no report of continuum
flux changes either. In this respect, the UV variability reported for
NGC\,4579 could be a further link between LINER nuclei and more
powerful AGN.  X-ray long term variability by significant factors
has been reported for M\,81 (Petre et al.\ 1993; Ishisaki et al.\
1996) as well as for NGC\,4579 (Serlemitsos, Ptak \& Yaqoob 1996).

\subsection{An advective-dominated accretion flow in NGC\,4395}

One possibility to explain the low luminosity for NGC\,4395 is the
presence of an advective-dominated accretion flow, or ADAF (e.g.,
Narayan \& Yi 1995; Narayan, Mahadevan \& Quataert 1998).  Such a disc
is optically thin and cooled by radial advection of heat rather than
radiation. Such discs differ from the optically thick variety is
several ways. ADAFs have a low radiative efficiency, are thermally and
viscously stable and have a vertical thickness similar to their radius
(i.e., the geometrically thin approximation does not hold).  The
combination of an outer, optically thick disk with an inner ADAF seems
capable of explaining some apparently under-luminous, hot black-hole
binary systems (e.g., Esin, McClintock \& Narayan 1997), although it
is unclear physically how to join the two parts of the disc. ADAFS
have also been proposed to explain under-luminous galactic nuclei,
most notably for Sagittarius A$^{*}$ (Narayan et al.\ 1998). For
NGC\,4395 an ADAF could allow for a `normal' mass black-hole while
maintaining consistency with the bolometric luminosity. However, our
estimation of the black-hole mass implies a radiative efficiency
higher than usually associated with ADAFs. For an upper limit for the
mass of the black-hole see also Ho (1998).

It has been claimed that dwarf AGN vary considerably less than
luminous Seyferts, both in the X-rays (Ptak et al.\ 1998) and the
optical (Shields \& Filippenko 1992; Ho, Filippenko \& Sargent 1996).
Ptak et al.\ (1998) propose that this is because dwarf AGN are ADAFs,
and have characteristically larger X-ray emitting regions. However in
NGC\,4395 we see clear large amplitude variability quite consistent
with more luminous Seyferts. Perhaps typical dwarf AGN have large
black holes together with low accretion rates and advection dominated
flows, whereas NGC\,4395 might have a small black hole and a higher
accretion rate (see section 4.5 and Ho 1998).

\subsection{Constraints on nuclear starburst models}

A plausible model for the IR,optical, UV and at least part of the
X-ray emission in low luminosity AGN, and perhaps all AGN, is the
starburst model developed by Terlevich and collaborators (see, e.g.,
Terlevich, Tenorio-Tagle \& Franco, 1992). NGC\,4395 offers a unique
opportunity to test this paradigm: with a blue nuclear absolute
magnitude of $\sim -10$ a single compact supernova remnant is enough
to account for its luminosity. We have significant constraints from
our spectroscopy, and from both short and long term variability.

In the starburst model, Type 1 AGN correspond to a phase in the
evolution of a metal rich star cluster with an age around 10-20 Myr,
where the luminosity is dominated by Type II supernovae which are
exploding within a dense medium.  These compact supernova remnants
(cSNR) evolve rapidly and so are highly luminous.  The energy output
is mostly in the extreme UV and X-ray region of the spectrum. The
optical continuum is a mixture of evolved stars and emission from the
SNRs, but fast moving fragments are also expected to produce the BLR
line emission. Indeed, the spectra of at least some luminous SN
exploding in HII regions have a striking resemblance to that of the
BLR of Seyfert galaxies (Filippenko 1989).  These cSNR in some well
documented cases radiate about $10^{51}$ ergs in under two years, thus
reaching peak luminosities well in excess of $10^{43}$ ergs s$^{-1}$
(Aretxaga et al.\ in preparation).

Our discovery of CaIIK absorption in NGC\,4395, without a corresponding
4000 \AA\ break, is strong evidence for a young stellar population, as
discussed in section 3.2.  However the observed variability rules out
the possibility that the blue light is simply young stars. It is
conceivable that the optical continuum is a mixture of early-type
stars and variable cSNR emission (or of course a mixture of early-type
stars and variable accretion disc emission).  As the CaIIK equivalent
width is a sensitive function of stellar type, the dilution required
to match our observed 1 \AA\ equivalent width depends on the age of
the cluster. It would be important to test the predictions of such
mixture models explicitly against the whole spectrum. We do not
attempt this here, but may do so in later work.

A significant constraint on starburst explanations of NGC\,4395 is the
long term stability. Although we have seen factor 2 variability on
relatively short time-scales, comparison with the Filippenko \&
Sargent (1989) observations, and indeed with the POSS image discussed
by Filippenko, Ho \& Sargent (1993), shows that there has been no
significant long term change in mean brightness beyond a factor of 2
in the last $\sim 40$ years. The secular variability in the starburst
model comes from the evolution of cSNRs. In more luminous Seyferts,
the superposition of cSNR events at a rate governed by the total mass
of the starburst could potentially explain the characteristic
month-time-scale flaring.

In the case of NGC 4395, we should be looking at the decline of a
single cSNR whose secular blue light-curve is given by $L_{B} \propto
t^{-11/7}$ (Aretxaga \& Terlevich 1994, Aretxaga, Cid Fernandes \&
Terlevich, 1997). The peak luminosity of the remnant will be a
function of the circumstellar density ($L_{B}^{peak} \propto
n^{3/4}$), with more luminous events ocurring in denser regions. We
note that the observed absolute magnitude of NGC\,4395 ($\sim -10$) is
much fainter than the absolute magnitude of a single cSNR near peak
($\sim -20$ and $-18.5$ for $n \sim 10^{7}$ and $10^{6}$ cm$^{-3}$,
respectively). This, together with the low rate of change seen in the
last 40 years, as mentioned above, suggests that we are observing the
late evolution of the cSNR. Using the $t^{-11/7}$ law we find that a
cSNR of age $\sim 300$ years old evolving in a medium with density $n
\sim 10^{7}$ cm$^{-3}$ would have shown a decline of 10 magnitudes
since maximum and of just 0.2 magnitudes in the last 40 years. On the
other hand, with a density of $10^{6}$ cm$^{-3}$ a $\sim 500$ year old
cSNR would have declined about 8.5 magnitudes since maximum and about
0.1 magnitudes in the last 40 years. Both cases are in agreement with
the observations.

We can also compare predicted line strengths by using the cSNR models
of Terlevich (1994).  If a significant fraction of the nuclear blue
luminosity is due to a starburst, then the predicted SN rate is $4.0
\times 10^{-5}$ yr$^{-1}$. We have seen that the nucleus has,

$L_{X} \sim 3 \times 10^{38}$ ergs s$^{-1}$

$L_{Bol} \sim 5 \times 10^{40}$ ergs s$^{-1}$

$L_{H\alpha} \sim 5.5 \times 10^{38}$ ergs s$^{-1}$

$L_{H\beta} \sim 8.5 \times 10^{37}$ ergs s$^{-1}$

\noindent
and the FWHM of the broad component of  the  broad  lines,  while
difficult to measure, is about 1000 Km/sec.

From table 1.1 and 1.2 of Terlevich (1994) it is possible to see that
this set of parameters corresponds to a cSNR evolving in a medium with
$n \sim 10^{7}$ cm$^{-3}$ and age equal or slightly larger than the
last entry corresponding to about 32 years \footnote{There is a
typographical error in table 1 of Terlevich 1994: the time for the
last entry is 31.5 years not 25.6 years}. The predicted values for 34
yrs are:

$L_{Bol} \sim 3.8 \times 10^{40}$ ergs s$^{-1}$

$L_{H\alpha} \sim 1.1 \times 10^{39}$ ergs s$^{-1}$

$L_{H\beta} \sim 2.5 \times 10^{38}$ ergs s$^{-1}$

\noindent
and the FWHM of the broad component about twice the shock velocity or
700 km s$^{-1}$. The line properties are then in rough agreement with
a cSNR of about 34 years old, somewhat younger than the age derived
from the absolute blue magnitude. Further work is needed to see if the
agreement can be improved.

We can also make an estimate of the expected X-ray fluxes of such a
cSNR.  According to the analytical approximations (Terlevich 1994),
for a cSNR evolving in a $n = 10^{7}$ CSMat(?) and t = 34 yr , the
leading shock should have an X-ray luminosity of $L_{x} \sim 3.8
\times 10^{40}$ ergs s$^{-1}$ in the $10^{-2} - 10$ keV band, and a
temperature of $T \sim 1. \times 10^{6}$ K. At a distance of 5.21 Mpc
and assuming a Raymond \& Smith model for Solar abundance with Log(T)
$= 6.0 $ and a hydrogen column of 21.15, the predicted count rate in
the {\it ROSAT\/} HRI band pass is $8 \times 10^{-4}$ counts s$^{-1}$
and in the {\it ROSAT\/} PSPC band pass would be $6 \times 10^{-3}$
counts s$^{-1}$ with hardness ratio H.R. $\sim 0.6$.  Thus we
conclude that the predicted fluxes for {\it ROSAT\/} are in good
agreement with our observations.

The short time-scale variability in the starburst model is produced by
cooling instabilities in the strongly radiative shock (Terlevich et
al.\ 1995; Plewa 1995) and also the result of fragments or density
fluctuations in the ejecta interacting with the outer thin shell (Cid
Fernandes et al.\ 1996). These mechanisms are capable of producing
strong variability in luminous AGN on time-scales from hours to
weeks. However it might be expected that such an old remnant would
have become relatively stable. It is currently not clear whether fast
optical and X-ray variability can be produced in such a case, and
further theoretical work is needed.

In summary, we have some evidence that a single cSNR could explain the
observed low luminosity optical and X-ray flux from NGC\,4395, if the
remnant is some tens to hundreds of years old. However, the observed
short term (days to months), large-amplitude optical and X-ray
variability, including changes in optical continuum shape, pose a
major challenge for such a model. Detailed modelling is required to
further test this hypothesis.

\section{Summary}

In this paper we present optical broad-band images and spectra as well
as {\it ROSAT\/} X-ray data for the nearest, most feeble known Seyfert
1 galaxy NGC\,4395.  The main results can be summarised as follows:

1. The optical continuum has been observed to vary by a factor $\sim
1.3$ at 6800 \AA\ and by a factor of $\sim 2.2$ at 3800 \AA\ in a
period of 6 months (between July 1996 and January 1997), becoming
bluer when brighter. A power law fit shows that the spectral shape
changed from an index $\alpha = 1.7$ to $\alpha = 1.1$.

2. One week of broad-band monitoring was obtained in June 1996. The B
and I bands show $\sim 20$ per cent variability in just 24 hours. The
inferred spectral shape is consistent with a power law fit with $\alpha
\sim 0$ and implies a change by a factor of $\sim 3$ at 4400 \AA\ in
just one month, when compared with the low state spectroscopy from
July 1996. This evidence shows that NGC\,4395 varies in much more
dramatic way than classic Seyfert galaxies.

4. UV spectroscopy shows a clear hump at $\sim 3000$ \AA\ consistent
with the small blue bump observed in classic AGN. The extrapolation of
the featureless UV continuum to higher frequencies assuming $\alpha =
1$ gives just enough ionizing photons to explain the observed
H$\alpha$ fluxes. However a deficit of photons is found when the
extrapolation is done assuming a steeper $\alpha = 1.7$ power law as
implied by the optical low state spectral shape. Anisotropic emission
from the central source and/or a different ionizing mechanism might
need to be considered to explain the lack of UV photons.

5. HRI and PSPC {\it ROSAT\/} data show that the flux from a weak
source consistent with the position of the nucleus changed by a factor
of 2 in 15 days. NGC\,4395 appears to be $\sim 26$ times less X-ray
loud than classic AGN.  The low luminosity can be explained as
substantial absorption along the line of sight. Spectral analysis of
the PSPC data is consistent with some intrinsic absorption.

6. Applying reverberation mapping results from classic AGN to
NGC\,4395 we infer a radius for the BLR of just $2 \times 10^{-4}$ pc!
This implies that the central source is emitting at $2 \times 10^{-3}
L_{Edd}$. However, these results were calculated by extrapolating the
observed properties of AGN $\ga 10^{4}$ times more luminous than
NGC\,4395 and, therefore, must be revisited when more observations on
very low luminosity AGN become available.

7. We report the discovery of a weak CaIIK absorption line (EW $\sim
1$ \AA), suggesting the presence of a young stellar cluster coincident
with the nucleus of NGC\,4395. The starlight component from the
cluster is estimated to be less than 30\% of the total flux at 4000
\AA. The cluster may be directly observed as a diffuse component in
{\it HST\/} optical imaging, suggesting a 10\% contribution. We expect
the IR CaII triplet to be detectable. Thus high resolution
spectroscopy can provide a direct measurement of the dynamical mass of
the young cluster.

8.  We show that NGC\,4395 can also potentially be explained by a
starburst with a single compact SNR of age 50--500 years producing the
broad lines, X-rays and the NLR.  The observed rapid variability is a
potential problem for this model. Although on theoretical grounds it
is expected that these radiatively dominated remnants to be highly
variable, it is unfortunate that no short term variability study is
available for known compact SNRs.

\section*{Acknowledgements}

PL acknowledges the support of a PPARC -- Fundaci\'on Andes research
studentship.  We thank Luis Ho and collaborators who kindly provided
us with their spectroscopic data of NGC\,4395. We also thank Elena
Terlevich for providing helpful comments of this work. The Jacobus
Kapteyn and Isaac Newton telescopes are operated on the island of La
Palma by the Royal Greenwich Observatory in the Spanish Observatorio
del Roque de los Muchachos of the Instituto de Astrof\'{\i}sica de
Canarias. The DSS plate used in this paper was based on photographic
data of the National Geographic Society -- Palomar Observatory Sky
Survey (NGS-POSS) obtained using the Oschin Telescope on Palomar
Mountain.  The NGS-POSS was funded by a grant from the National
Geographic Society to the California Institute of Technology.  The
plates were processed into the present compressed digital form with
their permission.  The Digitized Sky Survey was produced at the Space
Telescope Science Institute under US Government grant NAG W-2166.
Data were reduced using Starlink facilities. PL thanks Andrew Cooke
for his useful help throughout the writing of this paper.

{}

\end{document}